\documentclass[conference]{IEEEtran}

\ifCLASSINFOpdf
\else
\fi

\usepackage{tikz}
\usepackage{array}
\usepackage{tabularx} 
\usepackage{url}

\usepackage{acronym}
\usepackage{xspace}
\usepackage{tikz}
\usepackage{xfrac}
\usepackage{multirow}
\usepackage{makecell}
\usepackage{stfloats} %
\usepackage{float}
\usepackage{listings}
\usepackage{xcolor}
\usepackage{enumitem}
\usepackage{pifont}%
\usepackage{tabularx} %
\usepackage{mathtools}
\usepackage{balance}
\usepackage{paralist}
\usepackage{adjustbox}
\usepackage{booktabs}
\usepackage{hyperref}

\newcommand{\Sanctuary}{\textsc{Sanctuary}\xspace}
\newcommand{\Shelter}{\textsc{Shelter}\xspace}
\newcommand{\vSGX}{\textsc{vSGX}\xspace}
\newcommand{\NestedSGX}{\textsc{NestedSGX}\xspace}
\newcommand{\HyperEnclave}{\textsc{HyperEnclave}\xspace}
\newcommand{\Komodo}{\textsc{Komodo}\xspace}
\newcommand{\Tarnhelm}{\textsc{Tarnhelm}\xspace}
\newcommand{\Gramine}{\textsc{Gramine}\xspace}
\newcommand{\Occulum}{\textsc{Occulum}\xspace}
\newcommand{\BarriCCAde}{\textsc{BarriCCAde}\xspace}
\newcommand{\TrustShadow}{\textsc{TrustShadow}\xspace}

\newcommand{\Cage}{\textsc{Cage}\xspace}

\def\CPP{{C\nolinebreak[4]\hspace{-.05em}\raisebox{.35ex}{\footnotesize\bf ++}}\xspace}
\newcommand{\tool}[1]{\texttt{#1}\xspace}

\newcommand\Tstrut{\rule{0pt}{2.6ex}}         %
\newcommand\Bstrut{\rule[-0.9ex]{0pt}{0pt}}   %

\DeclareRobustCommand{\encircle}[1]{%
  \tikz[baseline=(X.base)] 
    \node (X) [draw, shape=circle, inner sep=0em,text width=1em, text centered] {\scriptsize #1};}

\newcolumntype{R}[2]{%
    >{\adjustbox{angle=#1,lap=\width-(#2)}\bgroup}%
    l%
    <{\egroup}%
}

\newcommand*{\fillcircle}[1]{%
  \raisebox{-0.15\height}{%
    \begin{tikzpicture}[scale=0.13]
      \draw (0,0) circle (1);
      \fill[black]
        (0,0) --
        (90:{#1>0?1:0})
        arc (90:90-#1*3.6:1) --
        cycle;
    \end{tikzpicture}%
  }%
}

\newcommand{\fullcircle}{\fillcircle{100}}
\newcommand{\halfcircle}{\fillcircle{50}}
\newcommand{\emptycircle}{\fillcircle{0}}

\begin{document}
\title{CCX: Enabling Unmodified Intel SGX Applications on Arm CCA}

\author{\IEEEauthorblockN{Matti Schulze}
	\IEEEauthorblockA{FAU Erlangen-Nürnberg \\ 
    matti.schulze@fau.de}
	\and
	\IEEEauthorblockN{Thorsten Holz}
	\IEEEauthorblockA{Max Planck Institute for Security and Privacy \\ 
    thorsten.holz@mpi-sp.org}
	\and
	\IEEEauthorblockN{Felix Freiling}
	\IEEEauthorblockA{FAU Erlangen-Nürnberg \\ 
    felix.freiling@fau.de}}

\maketitle

\begin{abstract}

    Novel confidential computing technologies such as Intel TDX, AMD SEV, and Arm CCA have recently emerged.
    In practice, due to its minimal trust boundaries, Intel SGX still remains widely used for enclave-based applications in cloud environments, including confidential cloud services, privacy-preserving communication, secure payment processing, and privacy-focused advertising.
    With the growing adoption of Arm CPUs in cloud systems, however, existing SGX applications face a significant portability challenge: they are tightly coupled to SGX-specific APIs and execution semantics.
    
    In this paper, we present the design and implementation of CCX, a framework that enables existing SGX applications to run on Arm CCA \emph{without} source code modification. 
    To this end, CCX redesigns SGX functionality within Arm CCA firmware, adapting SGX abstractions to CCA's architecture design while preserving full compatibility with existing applications originally developed for SGX. 
    We implemented a prototype of CCX on both the QEMU emulator and a Nitrogen8M development board.
    Our evaluation shows that CCX is capable of executing existing SGX applications without requiring source code changes, while providing security guarantees comparable to Intel SGX and achieving performance improvements in our evaluated settings.

\end{abstract}

\IEEEpeerreviewmaketitle

\section{Introduction}
\label{sec:intro}

Recurring vulnerabilities in system software, in particular operating systems (OSes) and hypervisors, have motivated the development of \emph{trusted execution environments}~(TEEs). 
TEEs execute code in hardware-isolated environments, protecting its confidentiality and integrity from potentially compromised system software. 
Early TEEs primarily targeted user-space applications through so-called \emph{enclaves}, which isolate a trusted component from the rest of the application.
For example, Intel SGX provides enclaves that enable fine-grained isolation of a trusted component (T-App) within an application from an untrusted component (U-App) \cite{sgx_prog_model}. 

However, in recent years, TEEs have increasingly targeted cloud computing, enabling isolation of entire virtual machines (\emph{confidential VMs}, or cVMs) even from the cloud provider's hypervisor.
Technologies supporting cVMs include AMD SEV~\cite{sev2020strengthening}, Intel TDX~\cite{DBLP:journals/csur/ChengOVAGJFB24}, and Arm CCA~\cite{DBLP:conf/seed/MulliganPSSV21,DBLP:conf/osdi/LiLDGNSS22}.

While cVMs enable flexible deployment of confidential workloads, they operate at VM granularity.
As a result, they have a larger trusted computing base (TCB) and lack support for fine-grained intra-process isolation~\cite{intel_sgx_tdx}
Therefore, Intel SGX~\cite{anati2013innovative,DBLP:conf/isca/HoekstraLPPC13,DBLP:conf/isca/McKeenABRSSS13}, while deprecated on Intel's desktop platforms~\cite{sgx_deprecated}, remains fully supported in Intel's server CPUs and is a popular choice in cloud environments.
Intel attributes this popularity to SGX's minimal trust boundary and precise execution model, which cVMs do not provide~\cite {intel_sgx_tdx}.
However, while developers who prioritize these qualities over the flexibility of cVMs can still choose to use SGX on Intel platforms, these properties are not available on other architectures.

On AMD platforms, this limitation has motivated several efforts to port SGX via virtualization~\cite{DBLP:conf/usenix/JiaLWCZYH22,DBLP:conf/sp/ZhaoLZL22,DBLP:conf/ndss/0001SMLZY0M025}. 
These approaches benefit from the shared x86-64 architecture of Intel and AMD CPUs, enabling the reuse of many underlying components and designs with minimal modifications.
However, because they rely on virtualization rather than hardware-enforced protection mechanisms, they cannot provide security guarantees equivalent to SGX or other hardware TEEs.
Furthermore, the ease of porting enabled by the shared x64 architecture does not carry over to other architectures, such as Arm CPUs, which differ significantly from x86-64 CPUs.
Although Arm CCA provides support for cVMs, it does not natively provide SGX-style user-space enclaves.
This issue is increasingly important as Arm CPUs are widely adopted in cloud environments~\cite{google_2024_arm_tau,aws_2022_arm,azure_2022_arm}.

Recent research has therefore explored how to provide enclave-like abstractions on Arm-based platforms.
For example, \Shelter~\cite{DBLP:conf/uss/ZhangHNZLHYH23} reuses Arm CCA hardware features to execute entire user-space applications in enclaves.
While this enables \emph{inter-process} isolation, such approaches cannot be compared with SGX, which offers \emph{intra-process} isolation.
This observation also motivated concurrent work, such as NanoZone~\cite{DBLP:journals/corr/abs-2506-07034} and HiveTEE~\cite{huang2026hivetee}, which repurpose CCA hardware primitives to implement intra-process isolation.

Although these approaches show promising results, their application in real-world settings raises two challenges.
First, SGX has a substantial application ecosystem.
It is widely used in practice, for example, by Microsoft~\cite{ms_payment}, IBM~\cite{german}, Signal~\cite{signal_contact,DBLP:conf/osdi/ConnellFSDP24}, Mozilla~\cite{mozilla}, Cosmian~\cite{cosmian}, or Flashbots~\cite{flashbots} as well as in academic research, as evidenced by a substantial body of literature using it~\cite{DBLP:journals/csur/WillM23}.
Second, the suggested programming models and SDKs of the new approaches are not as mature as SGX's programming model~\cite{sgx_prog_model} and the implementation offered by the SGX SDK~\cite{sgx_sdk}, which has evolved over many years. 
Existing SGX applications could, in principle, be ported to new Arm enclave frameworks, but doing so would require redesigning them for a different programming model, incurring substantial engineering cost and weakening the value of the existing SGX ecosystem.

This leaves a significant portability challenge:
Arm platforms are gaining traction in cloud environments~\cite{arm-marketshare}, while many confidential applications still rely on SGX's minimal trust boundaries afforded by intra-process isolation.
Existing approaches can provide enclave-like isolation on Arm, but they do not allow existing SGX applications to migrate transparently.
Instead, they require developers to adopt new and less mature APIs, SDKs, and execution models.

To address this problem, we present CCX, a framework that enables transparent migration of existing SGX applications to modern Arm systems. 
Like NanoZone~\cite{DBLP:journals/corr/abs-2506-07034} and HiveTEE~\cite{DBLP:journals/corr/abs-2506-07034,huang2026hivetee}, CCX repurposes Arm CCA for enclave-style isolation.
Unlike prior work, however, CCX focuses on SGX compatibility: we reimplement the SGX instruction interface within Arm CCA firmware and adapt SGX's architecture-dependent components (i.e., SDK, kernel module, and compiler) to Arm CCA.
This requires redesigning several SGX mechanisms, including enclave creation and entry, and memory management.
With this design, CCX preserves the SGX enclave abstraction on Arm systems \emph{without} requiring modifications of the source code.
Existing SGX applications only need to be recompiled with the CCX toolchain.
CCX effectively reproduces SGX’s enclave execution model on a fundamentally different ISA.
By preserving the semantics of SGX’s isolation features within a comparable threat model, CCX aims to provide security guarantees analogous to those of SGX. 
Hence, CCX relies on a minimal TCB that includes only the most privileged Arm firmware component, the Monitor.
As a result, CCX brings the security of SGX's mature intra-process isolation capabilities to Arm CPUs.

To validate our design in the absence of publicly available CCA hardware, we implemented CCX on the QEMU emulator and a Nitrogen8M development board.
Using these prototypes, we executed a diverse set of open-source SGX applications (e.g., a cryptographic library, a database, a federated learning framework, and a data analytics tool) without requiring \emph{any} source code modifications.
These results demonstrate that CCX can support realistic server-side SGX workloads on Arm systems.
Our performance evaluation further shows that CCX runs workloads at near-native speed and, in most evaluated cases, outperforms SGX.

\smallskip \noindent
In summary, we make the following key contributions:
\begin{compactitem}
  \item We present the design and implementation of CCX, a framework that enables existing SGX applications to run on Arm CCA without \emph{any} source code modifications. 
  \item We demonstrate the extensibility and practicality of CCX by adapting the SGX software stack, including the SDK, kernel module, and compiler, to Arm CCA while preserving compatibility with existing enclave applications.
  \item We build two prototypes of CCX on the QEMU emulator and on a Nitrogen8M development board. Based on these implementations, we conduct a comprehensive evaluation with multiple SGX applications and benchmarks. Our results show that CCX runs enclaves with near-native performance and effectively brings the benefits of SGX to modern Arm-based cloud environments.
\end{compactitem}

To foster further research and development, we open-source the full CCX framework.
We make the performance prototype available at \url{https://github.com/ma-schulze/ccx_nitrogen8m} and the functional prototype at \url{https://github.com/ma-schulze/ccx}.

\section{Background}
\label{sec:bg}
In the following, we provide an overview of the foundations of Intel SGX and Arm CCA. 
We first review SGX’s architecture and programming model, which CCX aims to reproduce on Arm platforms.
We then describe Arm CCA, whose primitives CCX leverages to realize SGX-like enclaves.

\subsection{Intel Software Guard Extensions (SGX)}

Intel SGX is a hardware extension available on most Intel server CPUs that enables secure execution of user-space enclaves, supporting fine-grained intra-process isolation.
First described in 2013~\cite{anati2013innovative,DBLP:conf/isca/HoekstraLPPC13,DBLP:conf/isca/McKeenABRSSS13}, processors supporting this extension became available starting in 2015.

\textbf{Hardware Architecture.} 
At the core of SGX lies the \emph{Enclave Page Cache} (EPC), a protected DRAM region secured by hardware encryption and strict access controls. 
Only dedicated SGX instructions can access the EPC, while all other accesses (e.g., from the OS, hypervisor, or firmware) trigger CPU faults. 
The suggested programming model for SGX uses these features to load security-critical user-space code into the EPC, creating so-called \emph{enclaves}. 

SGX management relies on two instruction classes: \texttt{ENCLU} for user-space and \texttt{ENCLS} for privileged operations. 
Both invoke microcode routines responsible for EPC management, with the leaf number in the \texttt{EAX} register selecting the specific routine (see Table~\ref{tab:sgx_instructions} in the appendix). 
SGX has evolved over time, introducing the SGX2 and AEX-Notify~\cite{DBLP:conf/uss/ConstableBCXXAK23} extensions that offer additional microcode routines. 
The available hardware configuration (e.g., EPC size or extension support) can be determined via \texttt{CPUID} calls.
This microcode-centric design is central to SGX’s execution model and poses a significant challenge for reproducing its semantics on other architectures.

Various data structures, mostly placed within the EPC, manage SGX resources. 
A key data structure is the \emph{SGX Enclave Control Structure} (SECS), which encodes an enclave's attributes, state, and identifier. 
Each enclave is associated with one or more \emph{Thread Control Structures} (TCS), which manage enclave threads by storing their execution context, including registers and instruction pointers. 
In addition, every EPC page is tracked by the \emph{Enclave Page Cache Map} (EPCM), which records its permissions, state, page type, SECS association, and virtual address. 
Together, these data structures govern the enclave lifecycle and enforce the isolation guarantees.

\begin{figure}[]
  \begin{center}
    \includegraphics[width=0.9\columnwidth]{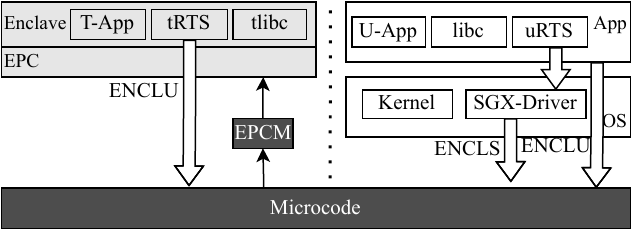}
  \end{center}
  \caption{Architecture of an SGX-App running on a typical SGX-enabled platform. Gray components are trusted, while white components are untrusted.}
  \label{fig:sgx} 
\end{figure}

\textbf{Programming Model.} 
Intel SGX suggests a specific enclave lifecycle and architecture, supported by the Intel SGX SDK~\cite{sgx_sdk} (see Figure~\ref{fig:sgx}). 
Applications are partitioned into an \emph{untrusted} component (U-App) and a \emph{trusted} component (T-App), the latter executing securely inside the EPC.
The enclave lifecycle begins when the U-App requests the kernel to create a new enclave via the \texttt{ECREATE} instruction. 
In this uninitialized state, enclave pages can be added with \texttt{EADD}, removed using \texttt{EREMOVE}, and the enclave's cryptographic measurement can be extended via \texttt{EEXTEND}. 
Once the setup is complete, \texttt{EINIT} transitions the enclave into an initialized state. 
From here, additional instructions support EPC management: pages may be blocked (\texttt{EBLOCK}), tracked (\texttt{ETRACK}), encrypted and saved to disk (\texttt{EWB}), and reintroduced into the enclave (\texttt{ELDU}/\texttt{ELDB}) to optimize EPC usage via page swapping. 
The model explicitly encourages writing pages to disk to avoid EPC exhaustion. 
However, this swapping, necessitated by the limited EPC size, can be a major performance bottleneck in SGX-based systems~\cite{DBLP:journals/pomacs/NgocBBTSFH19,hyperenclave_git}.

Enclaves are entered via \texttt{EENTER}, or resumed with \texttt{ERESUME} after an interrupt triggers an \emph{Asynchronous Enclave Exit} (AEX), during which control returns to the OS for interrupt handling. 
At runtime, an enclave can generate cryptographic keys (\texttt{EGETKEY}) and attestation reports (\texttt{EREPORT}).

SGX2 extends the model with dynamic page management, allowing an already-initialized enclave to add pages (\texttt{EAUG}), expand (\texttt{EMODPE}) or restrict (\texttt{EMODPR}) page permissions, and modify (\texttt{EMODPT}) page types. 
Such changes to an EPC page become active with the \texttt{EACCEPT} or \texttt{EACCEPTCOPY} instructions, the latter additionally copying an existing page to the targeted page in the EPC.

Finally, the AEX-Notify extension introduces an enclave-specific interrupt handler that executes after an AEX. 
This handler can implement mitigations against side-channel attacks, such as SGX-Step~\cite{DBLP:conf/sosp/BulckPS17}, by using the \texttt{EDECCSSA} instruction afterward to adjust the threads' internal TCS structure for resuming enclave execution.

\textbf{Intel SGX SDK.}
Most SGX applications use the Intel SGX SDK~\cite{sgx_sdk}, which provides untrusted libraries for the U-App and trusted ones for the T-App.
The untrusted runtime services library (uRTS) manages enclave creation and function execution by interfacing with the Linux kernel driver, which handles privileged \texttt{ENCLS} calls. 
On the enclave side, the trusted runtime services library (tRTS) implements core functionality, such as memory management via \texttt{ENCLU} calls, as well as a minimal C standard library (tlibc).

To simplify app development, untrusted and trusted code can reside within a single codebase. 
Intel's \emph{Enclave Definition Language} (EDL) specifies which parts of the code belong to the T-App. 
During preprocessing, the EDL generates the boilerplate for enclave entry calls (ECALLs) and outside calls (OCALLs), which are ultimately executed via the corresponding \texttt{ENCLU} instructions.
With this, SGX enables the isolation of code at a function granularity for intra-process isolation.
Due to the dependency of most SGX apps onto this SDK, any solution that aims to port SGX to other platforms must also port the SDK.
Since the SDK is tightly coupled to SGX’s microcode-based design, porting it requires redesigning low-level components while preserving API compatibility.

\subsection{Arm Confidential Compute Architecture}
\label{sec:bg:cca}

\begin{figure}
  \begin{center}
    \includegraphics[width=0.9\columnwidth]{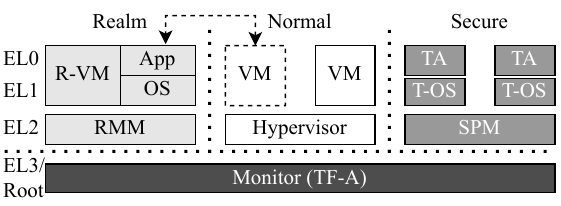}
  \end{center}
  \caption{Architecture of an Arm platform supporting Arm CCA (light gray components) and TrustZone (gray components). The Monitor manages the world separation. 
  }
  \label{fig:cca}
\end{figure}

Next, we outline the architecture of modern Arm CPUs, TrustZone, and Arm CCA, which Figure~\ref{fig:cca} illustrates.
Arm-A CPUs implement a vertical privilege model with four exception levels (EL3--EL0): EL3 hosts firmware, EL2 typically runs a hypervisor, EL1 the OS, and EL0 user applications. 
With Armv8.1, the \emph{Virtualization Host Extensions} (VHE) enable the OS to execute at EL2, while virtual machines run at EL1.
Most modern CPUs support VHE, making it the default for OSes like Linux. 
In general, higher ELs can access the resources of lower ELs and preempt their execution, while lower ELs have restricted access. 
Transitions to a higher EL occur via interrupts or dedicated instructions, such as \texttt{SVC} (user space to OS), \texttt{HVC} (VM to hypervisor), and \texttt{SMC} (OS or hypervisor to EL3 firmware). 
Additionally, some other instructions can also trigger synchronous exceptions to higher ELs, e.g., by triggering page faults, or if a higher EL disables the current EL's use of the feature associated with the instruction.

\textbf{Arm TrustZone.}
Introduced in 2004~\cite{DBLP:journals/csur/PintoS19}, Arm TrustZone extends the privilege model horizontally by introducing counterparts for EL2 to EL0 in the \emph{secure world} (SW), which EL3 is also assigned to. 
While commodity software (OS and hypervisor) runs in the \emph{normal world} (NW), the SW provides a hardware-protected TEE consisting of an isolated, trusted OS that executes trusted apps.
Arm Trusted Firmware (TF-A)~\cite{tfa} provides a reference implementation of the EL3 firmware, the \emph{Monitor}, which manages the TEE.

\textbf{Arm CCA.}
Introduced with Armv9 in 2021, CCA extends TrustZone to protect not only applications, but also entire VMs from malicious hypervisors~\cite{DBLP:conf/seed/MulliganPSSV21}.
However, it does not natively provide abstractions for inter- or intra-process isolation.”
CCA introduces the \emph{Realm Extension} (\texttt{FEAT\_RME}), which defines two additional security states: the \emph{realm world} for confidential VMs (cVMs) and the \emph{root world}, now hosting EL3. 
At realm EL2 (R-EL2), a new firmware component, the \emph{Realm Management Monitor} (RMM), secures the execution of cVMs, while the hypervisor continues to manage scheduling and resource allocation.
A new hardware component, the \emph{Granule Protection Table} (GPT), enforces memory isolation by associating each physical page with a security state. 
The GPT is a dynamically updatable structure similar to memory translation tables and can only be written by the Monitor. 
While there is usually a single GPT, its core-local registers allow each core to be assigned a distinct GPT if required.

Table~\ref{tab:cca_pas} summarizes the access permissions for CCA's security states. Only software executing in the realm of a secure state may access its respective memory regions, while EL3 has universal access.
Pages may also be marked as entirely inaccessible. 
Any unauthorized access attempt results in a \emph{Granule Protection Fault}~(GPF).

\begin{table}
  \caption{Overview of Arm CCA memory access permissions.} 
  \label{tab:cca_pas}
  \centering
  \resizebox{0.8\columnwidth}{!}{
    \begin{tabularx}{\columnwidth}{Xcccc}
      \hline
      \textbf{Security State} & \textbf{Normal} & \textbf{Secure} & \textbf{Realm} & \textbf{Root} \\ \hline
      \textbf{Normal}         & Y  & N  &  N    & N \\
      \textbf{Secure}         & Y  & Y  &  N    & N \\
      \textbf{Realm}          & Y  & N  &  Y    & N \\
      \textbf{Root}           & Y  & Y  &  Y    & Y \\
      \hline
    \end{tabularx}}
\end{table}

\section{Threat Model}
\label{sec:threat_model} 

Throughout the rest of this paper, we assume a threat model largely following the default Arm CCA assumptions.
In this model, the EL3 firmware is trusted, while all other software components in the normal, secure, and realm world (including other enclaves) are considered untrusted and potentially malicious.
This explicitly includes the RMM, which we also consider untrusted.
This is feasible because CCX does not rely on the RMM to enforce isolation; instead, it builds directly on hardware-enforced protections managed by the Monitor.
This reduces the TCB to the Monitor alone, which cannot be excluded because it controls the GPT that enforces enclave isolation.
With this, we trust only the minimal number of components possible on Arm systems, akin to SGX, which also only requires trusting the microcode.
We assume that the hardware behaves correctly and that neither the EL3 Monitor nor the target CCX enclave contains vulnerabilities.

Assuming control over the untrusted components, an attacker may try to extract or tamper with enclave data.
CCX, like Intel SGX~\cite{DBLP:journals/iacr/CostanD16} and Arm CCA~\cite{arm_cca_sec_model}, ensures confidentiality and integrity of the enclave, but does not guarantee availability.
The CCA threat model considers software-based side-channel attacks, including timing and Rowhammer-style attacks~\cite {10.5555/3241094.3241097,10.5555/3241094.3241096,rowhammer}.
However, similar to SGX and consistent with prior work~\cite{DBLP:conf/uss/ZhangHNZLHYH23,DBLP:journals/corr/abs-2506-07034},
CCX does not aim to fully mitigate such side channels and inherits the side-channel characteristics of the underlying platform.
Similarly, while considered in the CCAs threat model, we exclude hardware-level attacks.
Under this threat model, CCX aims to provide security guarantees comparable to those of SGX enclaves on Intel platforms.

\section{Design}
\label{sec:design}

We design CCX to tackle the challenge of executing SGX enclaves on Arm platforms, thereby bringing intra-process isolation built on a mature ecosystem to Arm systems.
Its design is guided by three main objectives. 
First, CCX must support the execution of existing SGX applications on Arm systems \emph{without} requiring any source code modifications. 
This guarantees that the wide range of existing SGX apps can then be used on top of CCX without requiring the significant developer effort needed by other solutions.
Second, it should deliver performance comparable to, or better than, native SGX to ensure practical deployment. 
Finally, CCX must maintain security guarantees comparable to those of SGX, preserving confidentiality and integrity under a strong adversarial model.
In the following, we describe the design principles and architecture of CCX that enable these goals.

\subsection{Design Considerations}
To achieve these goals, we first discuss design options from prior work aiming to bring SGX to other x64-based platforms~\cite{DBLP:conf/sp/ZhaoLZL22,DBLP:conf/usenix/JiaLWCZYH22, DBLP:conf/ndss/0001SMLZY0M025}.
At the core of these existing approaches lies a transparent replacement for SGX's microinstructions that preserves or strengthens its security guarantees.
For CCX, we refer to this component as the \emph{CCX-Core}.
It intercepts all \texttt{ENCLS} and \texttt{ENCLU} instructions and manages isolated memory (i.e., the EPC) independently of the host.
Crucially, it must be protected from all untrusted software components.

\HyperEnclave~\cite{DBLP:conf/usenix/JiaLWCZYH22} places this component in the hypervisor layer.
However, on Arm systems with VHE, the host OS typically executes at EL2, effectively eliminating a privileged layer that could isolate the CCX-Core without invasive changes to the system architecture.
Moreover, relying solely on virtualization weakens security compared to hardware-enforced isolation.
In contrast, \vSGX~\cite{DBLP:conf/sp/ZhaoLZL22} executes enclaves inside a cVM on top of an OS that forwards communication between trusted and untrusted components.
This approach increases the TCB and introduces significant performance overhead due to frequent cross-boundary interactions.

A CCA-specific alternative would be to place the CCX-Core inside the RMM.
However, all communication between the NW and the RMM must be mediated by EL3, introducing additional context switches.
Given the frequent interactions between U-App and T-App in SGX, this leads to both performance overhead and an expanded TCB, as the RMM would need to be trusted.

\NestedSGX~\cite{DBLP:conf/ndss/0001SMLZY0M025} leverages AMD virtual machine privilege levels~(VMPLs) to enable enclaves within a cVM.
Although CCA introduces a similar concept with \emph{planes}~\cite{rmm_spec}, such approaches only allow enclaves to be created within guests.
This contradicts SGX’s model, where enclaves are created by the host, and would again require trusting the RMM.

Instead, CCX adopts a different approach.
On Arm systems, EL3 firmware manages security-critical functionality, analogous in some respects to microcode on x64.
We therefore implement the CCX-Core at EL3, leveraging its privileged position to enforce isolation while minimizing the TCB.
Since no other software component can access EL3 resources, this placement establishes a strong and well-defined trust boundary.

\subsection{Design Overview}

\begin{figure}[tb]
  \begin{center}
    \includegraphics[width=0.95\columnwidth]{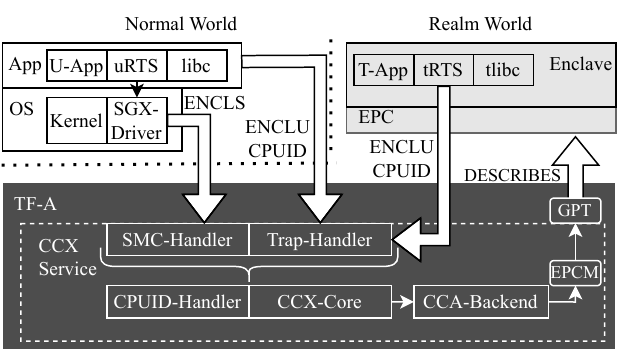}
  \end{center}
  \caption{Overview of CCX. Trusted components are gray; untrusted components are white. 
  The CCX-Service primarily includes the microprogram emulator provided by the CCX-Core and supporting components operating within the TF-A. }
  \label{fig:ccx}
\end{figure}

Due to these considerations, we integrate CCX into the EL3 Monitor implemented by TF-A.
Figure~\ref{fig:ccx} provides a high-level overview of the design.
In general, all enclave microinstructions executed by the commodity OS or a U-App running in the NW are routed to EL3, where the CCX-Core handles them. 
To compile existing SGX apps for Armv9 without source code modifications or loss of functionality, we also develop the \emph{CCX-toolchain}, a modified version of the SGX toolchain that includes the kernel module, SGX SDK, and compiler. 
These components, particularly the SGX SDK, require extensive modifications due to architecture-specific elements, as discussed in Section~\ref{sec:implementation}.
By using the existing \texttt{ENCLS} calls, the SGX kernel module and the U-App can invoke the CCX-Core to place the T-App in the realm world, enabling full CCA security guarantees. 
The U-App and T-App can then interact with each other and the CCX-Core via \texttt{ENCLU}, using the same semantics of the original SGX instructions. 
This can be used to call trusted or untrusted functions (ECALLs and OCALLs), while the T-App can also interact with the CCX-Core, e.g., to generate cryptographic attestation reports (via \texttt{EREPORT}) or cryptographic keys for sealed storage (via \texttt{EGETKEY}).

The key to enabling all of this is a transparent interface that forwards the execution of these instructions to the CCX-Core. 
This interface is implemented by various CCX components, together with the CCX-Core, forming the \emph{CCX-Service}.

\subsection{CCX-Service}

To implement SGX instructions in CCX, we reroute them to the appropriate handlers.
Specifically, \texttt{ENCLS} calls are intercepted by the \emph{SMC-Handler}, and \texttt{ENCLU} calls are handled by the \emph{Trap-Handler}.
Similarly, \texttt{CPUID} instructions are emulated by the \emph{CPUID-Handler}.
Both the SMC and Trap handlers execute instructions from the \emph{SGX microprogram table}, which is invoked via the corresponding \texttt{ENCLS} and \texttt{ENCLU} calls.

Since \texttt{ENCLS} instructions are executed from the OS (i.e., EL1 or EL2), we can simply replace them with the \texttt{SMC} instruction caught by the SMC-Handler. 
However, \texttt{ENCLU} and \texttt{CPUID} calls must be directly accessible from EL0 and directly trap execution to EL3, which no instruction directly implements in the current Armv9 ISA. 
Using a trampoline in the OS introduces context-switch overhead and breaks transparency, potentially violating our performance and functionality goals. 
Thus, alternatives to directly trap from EL0 to EL3 are crucial. 

\textbf{Trap-Handler.} 
Achieving synchronous trapping from EL0 to EL3 requires instructions that satisfy three core criteria: 
\begin{compactenum}
  \item Execution may result in a synchronous EL3 trap.
  \item Conditions for these traps can be reliably configured.
  \item Another EL cannot intercept the trap to EL3.
\end{compactenum}
The ideal instruction relates to security-critical features, which can generally be disabled from EL3 and, in this case, naturally generate synchronous EL3 traps when executed from EL0. 
Exemplary candidates for this include instructions that access hardware registers associated with such features. 
Accessing these from EL0 then yields a synchronous abort to EL3, which CCX can subsequently interpret as an \texttt{ENCLU} call. 
The aarch64 general-purpose registers \texttt{x0} to \texttt{x3} substitute x64's \texttt{EAX}, \texttt{EBX}, \texttt{ECX}, \texttt{EDX} for arguments.
Section~\ref{sec:implementation} details the instruction selected for CCX. 

\textbf{CPUID-Handler.}
By directly trapping to EL3, we can also emulate \texttt{CPUID} instructions at EL3. 
The CPUID-Handler executes different operations depending on the requested leaf function. 
\texttt{EAX} and \texttt{ECX} (or \texttt{x1} and \texttt{x3} in CCX) define the executed leaf and sub-leaf.
The leaf functions \texttt{0x12} and its sub-leafs return information about the SGX configuration, including EPC location and size. 

\begin{figure}[tb]
  \begin{center}
    \includegraphics[width=0.9\columnwidth]{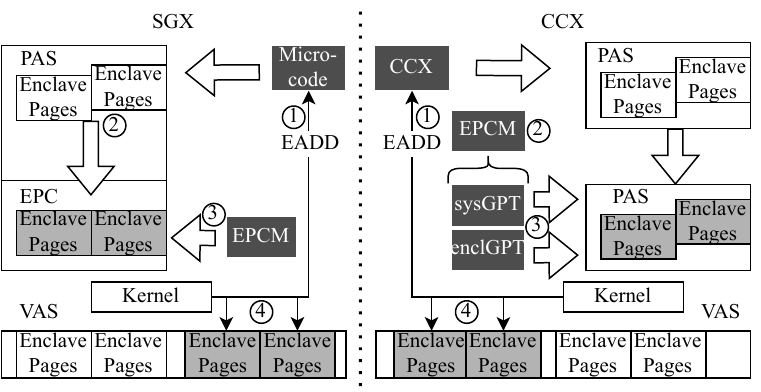}
  \end{center}
  \caption{Overview of different approaches to enclave memory management of SGX and CCX. Gray components are trusted, white components are untrusted, and dark gray components are trusted hardware and firmware. }
  \label{fig:memory_management}
\end{figure}

\textbf{CCX-Core.}
The CCX-Core module emulates SGX microprogram instructions whenever an \texttt{ENCLU} or \texttt{ENCLS} instruction triggers a trap. 
Each trap is identified by a leaf number, which determines the corresponding microprogram to execute.
While SGX implicitly manages side effects such as CPU state changes or hardware interactions, replicating these behaviors on Arm platforms requires additional handling within CCX.
For example, while some SGX instructions can be emulated directly, others, particularly those modifying CPU state (\texttt{EENTER}/\texttt{ERESUME} and \texttt{EEXIT}) or enclave memory state (\texttt{EADD}, \texttt{EREMOVE}, \texttt{EAUG}, \texttt{ELDU/B}, and \texttt{EWB}), require significant adaptation for Arm platforms. 
Hence, each CCX microcode instruction can invoke its corresponding functionality in the \emph{CCA-Backend} after executing the SGX microprogram.
Figure~\ref{fig:memory_management} and the following discussion highlight key differences in memory management between SGX and CCX, particularly in EPC management. 

\textbf{Memory Management.}
In SGX (left side of Figure~\ref{fig:memory_management}), the EPC is a fixed main memory segment. 
The kernel adds a page to an enclave in the EPC using the \texttt{EADD} instruction~\encircle{1}; microcode copies it into the EPC~\encircle{2} and updates the EPCM~\encircle{3}, assigning the EPC page to the enclave. 
The kernel then maps the new page to a new virtual address for the enclave~\encircle{4}. 
SGX's \emph{Page Miss Handler} (PMH) ensures that only the owning enclave can access this EPC page.

CCX takes a different approach (right side of Figure~\ref{fig:memory_management}).
The EPC resides in realm memory, while the EPCM is placed in root world memory, using the GPT for world assignment. 
When the kernel requests to add a page~\encircle{1}, it is added to the EPCM~\encircle{2}.
Our design introduces several improvements over SGX that address limitations arising from the EPC's static nature. 
In SGX, given its fixed size, only a limited number of EPC pages can be loaded at once.
When it becomes full, SGX must swap pages to and from disk using \texttt{EWB} and \texttt{ELDB/U}.
This page swapping is expensive, potentially consuming hundreds of thousands of cycles, and represents a major performance bottleneck when the EPC is full~\cite{DBLP:journals/pomacs/NgocBBTSFH19}. 
In memory-intensive scenarios, this can reduce SGX performance by up to 50\%~\cite{DBLP:conf/usenix/JiaLWCZYH22}. 
In contrast, the dynamic nature of the GPT allows pages to be mapped on demand, expanding the EPC without requiring \texttt{EWB} and \texttt{ELDB/U} instructions. 
In CCX, pages are hence dynamically marked as realm pages in the GPT when added to an enclave. 
In Section~\ref{sec:eval}, we evaluate the resulting performance improvements. 

Since all enclaves run in the realm world, this could enable cross-enclave attacks, where one enclave maps and accesses the memory of another enclave, which the PMH prevents for SGX~\cite{DBLP:journals/iacr/CostanD16}.
To prevent such attacks in CCX, we adopt a multi-GPT approach, inspired by \Shelter~\cite{DBLP:conf/uss/ZhangHNZLHYH23}. 
In addition to the system-wide GPT (sysGPT), the CCX-Core generates a dedicated GPT (enclGPT) upon creating a new enclave. 
Adding a page marks it as realm memory in enclGPT and as inaccessible in every other GPT, including sysGPT~\encircle{3}.
Afterward, the kernel adds new memory mappings for the enclave page to the app~\encircle{4}. 
This design enables another improvement: dynamic in-place EPC assignment eliminates the need to copy pages when creating enclaves, thereby improving enclave creation performance.
Since the U-App expects pages to be mapped to a new virtual address after being added to an enclave (due to the copying), CCX instead double-maps the new EPC page's physical address to mimic the original behavior.

To use the multi-GPT approach, enclave entries and exits are monitored.
Since the T-App operates in R-EL0, such entries and exits require context switches between the realm world and the realm.
Upon an enclave entry, the CPU core's GPT is switched to the enclave's GPT, returning to the system GPT on exit. 
Thus, no untrusted software component, including the RMM (which uses the sysGPT), can access enclave memory. 
In Section~\ref{sec:sec_discussion}, we discuss the security implications of this multi-GPT approach in detail.

\textbf{Attestation.}
In Intel SGX, the attestation process is based upon the instructions \texttt{EREPORT} and \texttt{EGETKEY}. 
For local attestation, two enclaves can attest each other by exclusively using these primitives, without involving any other party.
Remote attestation requires additional support from privileged enclaves, the \emph{Provisioning Enclave}~(PvE) and the \emph{Quoting Enclave}~(QE). 
The PvE uses a privileged variant of \texttt{EGETKEY} to retrieve the platform's \emph{Provisioning Key}~(PK), which is used in the provisioning process against Intel's provisioning servers to obtain the platform's \emph{Attestation Key}~(AK). 
The QE uses this key to sign reports from other enclaves, generating \emph{quotes} which can be verified using Intel's attestation services~\cite{DBLP:journals/iacr/CostanD16}. 

Since no CCA-enabled hardware is available at the time of writing, CCX does not have access to valid keys to identify its platform. 
Therefore, it is currently also impossible to implement the remote infrastructure needed to verify platforms that generate AKs, making proper remote attestation at the point of writing impossible.
Therefore, while CCX implements the necessary firmware functions to retrieve privileged keys, it is still--for now--only feasible to implement local attestation, a limitation also recognized in the related literature~\cite{DBLP:conf/uss/ZhangHNZLHYH23}.
Nonetheless, we implement the local attestation architecture fully and evaluate it in Section~\ref{sec:eval}.
Furthermore, we present a theoretical framework for implementing remote attestation for CCX once actual CCA hardware becomes available.

For CCX, the platform could be identified using the immutable \emph{CCA Platform Attestation Key} (CPAK), issued by the platform provider and implemented in hardware (e.g., eFuses)~\cite{arm_cca_sec_model}.
This key, together with the boot measurements, can be used to derive a PK, which the PvE can then retrieve.
This PK can then be sent to the provisioning service as proof of ownership to retrieve the platform's AK.
With this key, the remainder of the remote attestation process can be executed as in SGX.
First, the AK is sealed by the PvE using the \emph{Provisioning Seal Key} (PSK), which is also retrieved via \texttt{EGETKEY}. 
For remote attestation, an enclave can send its report (retrieved via \texttt{EREPORT}) to the QE, which unseals the stored AK using the PSK, signs the report with it, and forwards the result to the calling enclave, thus completing the remote attestation process.

\section{Implementation}
\label{sec:implementation}
We describe the implementation of the CCX-Service and the CCX-Toolchain.
In total, CCX comprises approximately 4,400 SLOC in EL3 Monitor, 6,000 SLOC for the SGX SDK port, and 650 SLOC of kernel modifications.

\subsection{SGX-Service}

\textbf{Trap- and SMC-Handler.}
To implement CCX's \texttt{ENCLU} gadget, we require a register whose access from EL0 can be configured to trap to EL3.
We select the PMU register \texttt{PMCCNTR\_EL0}, as user-space access to the PMU is uncommon or even undesirable.
Section~\ref{sec:discussion} further discusses the implications of this. 
By configuring \texttt{MDCR\_EL3.TPM} and \texttt{PMUSERENR\_EL0.EN}, accesses from EL0 trigger synchronous traps to EL3.

Although lower ELs can intercept these accesses, this is consistent with SGX, where the OS can interrupt \texttt{ENCLU} execution (e.g., via breakpoints).
Thus, requiring OS cooperation does not weaken the threat model, as availability is not guaranteed in SGX either.

\newcounter{foo}
\renewcommand{\figurename}{Listing.}
\setcounter{foo}{\value{figure}}
\setcounter{figure}{0}
\begin{figure}
  \centering
  \caption{ENCLU gadget for CCX.}
  \begin{minipage}{.9\columnwidth}
    \begin{lstlisting}[label=lst:enclu, frame=single, basicstyle=\ttfamily\footnotesize, breaklines=true]
mov x0, SMC_ID      // the CCX identifier
mov x1, LEAF        // SGX/CPUID leaf (EAX)
mov x2, arg1        // first argument(RBX)
mov x3, arg2        // second argument (RCX)
mov x4, arg3        // third argument (RDX)
msr pmccntr_el0, x0 // trigger ENCLU
    \end{lstlisting}
  \end{minipage}
\end{figure}
\renewcommand{\figurename}{Fig.}
\setcounter{figure}{\value{foo}}

Enclaves executing in R-EL0 use the same mechanism to implement \texttt{ENCLU}.
To prevent interception by the RMM, we disable PMU trapping at EL2 by clearing \texttt{MDCR\_EL2.TPM} and, if supported, \texttt{HDFGRTR\_EL2.PMCCNTR\_EL0}.
Since these registers are core-local, configuring them before enclave entry prevents the RMM from modifying them at runtime.
With this, \texttt{ENCLU} calls of an enclave cannot be intercepted by the RMM and are always routed to El3, which is important since the RMM is untrusted.

Once the registers are configured, we can use the gadget shown in Listing~\ref{lst:enclu} to trap execution from EL0 to EL3 atomically. 
The CCX toolchain is adapted to invoke this gadget in place of \texttt{ENCLU} or \texttt{CPUID}. 
For \texttt{ENCLS} calls originating from a privileged context (i.e., EL2 or EL1), the same gadget could be used. 
However, from these EL we can also use the default kernel SMC interface to emulate both \texttt{ENCLS} and \texttt{CPUID} calls, which CCX does for simplicity.

\textbf{CCX-Core.} 
The CCX-Core executes SGX microprograms upon receiving \texttt{ENCLS} or \texttt{ENCLU} traps.
Since EL3 cannot be interrupted by lower ELs, execution is atomic from their perspective.
SGX microprograms perform various cryptographic operations, such as \texttt{EINIT} for enclave signature verification and measurement validation, \texttt{EWB}/\texttt{ELDB/U} for encrypting swapped pages, and \texttt{EREPORT}/\texttt{EGETKEY} to generate cryptographic reports and keys. 
In CCX, we implement these functions using the \emph{mbed~TLS} library~\cite{mbedtls}, which TF-A~\cite{tfa} also uses by default.

Furthermore, some SGX data structures require adaptation due to architectural differences.
For example, the \texttt{SECS} structure must reflect Arm-specific context management, while \texttt{TCS} is adapted to Arm’s thread-local storage model (e.g., \texttt{TPIDR\_EL0} instead of segmentation registers).

SGX enclaves transition execution to the OS upon receiving interrupts. 
With Arm CCA, this would normally involve forwarding execution to RMM, then to the NW OS (routed through EL3). 
However, to maintain a minimal trust boundary, we do not trust the RMM (see Section~\ref{sec:threat_model}). 
Instead, we leverage the firmware to direct all external interrupts to EL3 using the \texttt{SCR\_EL3} register. 
Since this register always takes priority over any interrupt routing defined by lower ELs, the RMM cannot interfere with this.
Internally triggered exceptions (e.g., page faults) generally cannot be routed to EL3. 
To facilitate the switch without trusting the RMM, each enclave includes a trampoline page in its text segment that contains exception handlers. 
The exception table base register (\texttt{VBAR\_EL1}) is set to this page when entering the enclave. 
Each exception handler on this page executes a switch to CCX in EL3, which then forwards the exceptions to the NW OS.
The NW OS exception link register is also set to an entry point in the U-App, where an \texttt{ERESUME} is executed to continue the enclave execution afterward.  
Furthermore, as described above, all relevant EL2 registers are cleared before enclave entry, and since they are core-local, they cannot be modified by the RMM at runtime.
With this, exceptions cannot be routed to the RMM instead and are always taken to EL3.

\textbf{CCA-Backend.} 
With the EPC residing in realm memory, protected pages are accessed and executed there. 
Enclave entry via \texttt{EENTER} or \texttt{ERESUME} requires setting up a corresponding exception return context, which is handled by the CCA-Backend.
To do this, we use TF-A's built-in functionality to create a new CPU context on the first entry of a TCS and save it to its state save area. 
This context is then reused for each subsequent entry using this TCS.

Note that unlike Arm, SGX maintains a unified CPU state across enclave transitions.
To emulate this, CCX copies relevant registers (e.g., general-purpose and memory-management registers) into the enclave context on entry and restores them on exit, preserving SGX semantics.

\subsection{CCX-Toolchain}
Most SGX applications rely on the SGX SDK, which is tailored for x64 CPUs and contains extensive platform-specific code.
Adapting the SDK for CCX is hence necessary. 
Although this requires substantial engineering effort, the details are beyond the scope of this paper.
In brief, we replace SDK enclave \texttt{ENCLU} calls with our custom gadget and adjust all assembly code in the SDK for Arm. 
Instances where the SDK handles Intel-specific behavior (e.g., thread-local storage or context save states) are revised to reflect the specifics of the Arm architecture.
To assess the effort required to implement these adaptations, we compute the SLOC modified in the SGX SDK using \texttt{diff}, which amounts to approximately 6,000 new lines of code.
The SGX kernel module, which is part of the CCX-Toolchain, must reflect these adaptations. 
Here, we evaluate these changes to approximately 650 SLOC.
SGX applications may also directly invoke the wrappers for \texttt{ENCLS} and \texttt{ENCLU} calls. 
To support this, we extend GCC to provide the corresponding gadgets when compiling for Arm targets. These wrappers internally use the custom gadget to perform the necessary calls.

\section{Evaluation}
\label{sec:eval}
We now evaluate how CCX meets the goals defined in Section~\ref{sec:design}. 
First, we assess its functional correctness by verifying adherence to the SGX specification and by testing real-world applications.
Next, we analyze performance using micro- and macro-benchmarks.
Finally, we evaluate the security impact, focusing on the TCB.

\subsection{Goal~1: Functionality}
\begin{table}
\caption{Sample SGX SDK apps ported to CCX.}
\centering
  \resizebox{0.8\columnwidth}{!}{
\begin{tabularx}{\columnwidth}{ lX }
\hline \Tstrut 
\textbf{Name}                     & \textbf{Functionality} \Bstrut                                                                 \\ \hline \Tstrut
SampleEnclave                     & Contains test cases for ECALL/OCALL functionalities                                     \\
CXX17SGXDemo                      & Contains test cases for modern C++ runtime features in the enclave  \\
LocalAttestation                  & Local attestation between enclaves                                         \\
SealUnseal                        & Uses the enclave's sealing keys to secure, store, and retrieve data                       \\
SampleAEXNotify                   & Test case for AEX-Notify                         \Bstrut \\  \hline 
\end{tabularx}
}
\label{tab:eval_correctness}
\end{table}

\noindent
\textbf{Functional Correctness.}
At the time of writing, no Arm CCA hardware is publicly available.
Therefore, to evaluate the functional correctness of CCX, we build a feature-complete prototype for the QEMU emulator (version 9.2.3) with support for Arm CCA, running a full CCA firmware stack (TF-A version 2.10.0 and RMM version 0.4.0), and Linux (version 6.2.0).
In this environment, we can then execute the sample applications of the Intel SGX SDK to assess CCX's functional correctness.
These apps demonstrate typical SGX enclave behaviors and cover the different enclave functions. 
We select a representative subset that covers all SGX microcode functionalities, listed in Table~\ref{tab:eval_correctness}. 
\tool{SampleEnclave} contains a list of ECALLs and OCALLs, testing the correctness of parameter and return value forwarding on enclave entry and exits, multithreading, and the SGX2-specific features.
\tool{CXX17SGXDemo} tests whether the enclave runtime provided by the SGX-SDK correctly implements the \CPP runtime.
Next, \tool{LocalAttestation} checks a local attestation setup between two enclaves, each generating attestation reports to be verified by the other enclave. 
When the attestation is successful, the enclaves use the SGX-SDK's crypto library (based on OpenSSL) to establish a secure channel to exchange messages. 
\tool{SealUnseal} validates sealing and unsealing by having two enclaves exchange sealed data.
Finally, \tool{SampleAEXNotify} checks whether AEX-Notify is implemented correctly.

We cross-compiled these apps using the CCX toolchain without any source code modifications. 
Upon execution, they showed \emph{no} difference in behavior compared to their SGX versions, indicating the correctness of our approach.

\textbf{Real-World Applicability.}
To evaluate CCX on real-world server-side workloads, we use it to deploy several open-source SGX enclaves on Arm.
Table~\ref{tab:eval_rw} lists the chosen apps and describes their functionality.
\tool{SGX-kmeans} runs the k-means algorithm on a set of points securely inside an enclave. 
\tool{TaLoS} is a patch of LibreSSL, moving all security-critical cryptographic operations to an SGX enclave~\cite{talos}. 
The resulting library can then be used as a drop-in replacement for the default LibreSSL library. 
We deploy \tool{TaLoS} with CCX by using it as the SSL library of \tool{nginx}~\cite{nginx}, which was, in turn, used to host a web server. 
Next, \tool{SGX\_SQLite} is a port of \tool{SQLite} that protects its database memory with an SGX enclave. 
\tool{TrustFL}~\cite{DBLP:conf/infocom/ZhangLZLWW20} is a trusted federated learning framework whose test cases we executed.
Lastly, \tool{secure-analytics-sgx} performs privacy-preserving analytics in an SGX enclave~\cite{DBLP:conf/esorics/ChandraKLKKT17}. 

We successfully ported, executed, and tested all applications using CCX. 
We also run the same tests on actual SGX hardware and compare the results, which show no differences in behavior.
Again, this experiment demonstrates CCX's functional correctness for real-world workloads.

\begin{table}
\caption{Open-source SGX apps ported to Arm via CCX.}
\centering
  \resizebox{.8\columnwidth}{!}{
\begin{tabularx}{\columnwidth}{ lX }
\hline \Tstrut 
\textbf{Name}                                         & \textbf{Functionality} \Bstrut                                          \\ \hline \Tstrut
SGX-kmeans~\cite{sgx_kmeans}                          & Secure kmeans clustering                                        \\
  TaLoS~\cite{talos}                                  & LibreSSL patch using SGX to protect cryptography           \\
  SGX\_SQLite~\cite{sgxsqlite}                        & SGX protected SQLite                                  \\
  TrustFL~\cite{DBLP:conf/infocom/ZhangLZLWW20}       & Secure federated learning                                                \\  
  secure-analytics-sgx~\cite{DBLP:conf/esorics/ChandraKLKKT17}               & Secure Data Analytics \Bstrut \\  \hline 
\end{tabularx}
}
\label{tab:eval_rw}
\end{table}

\subsection{Goal~2: Performance}
\begin{figure*}
  \begin{center}
    \includegraphics[width=0.9\linewidth]{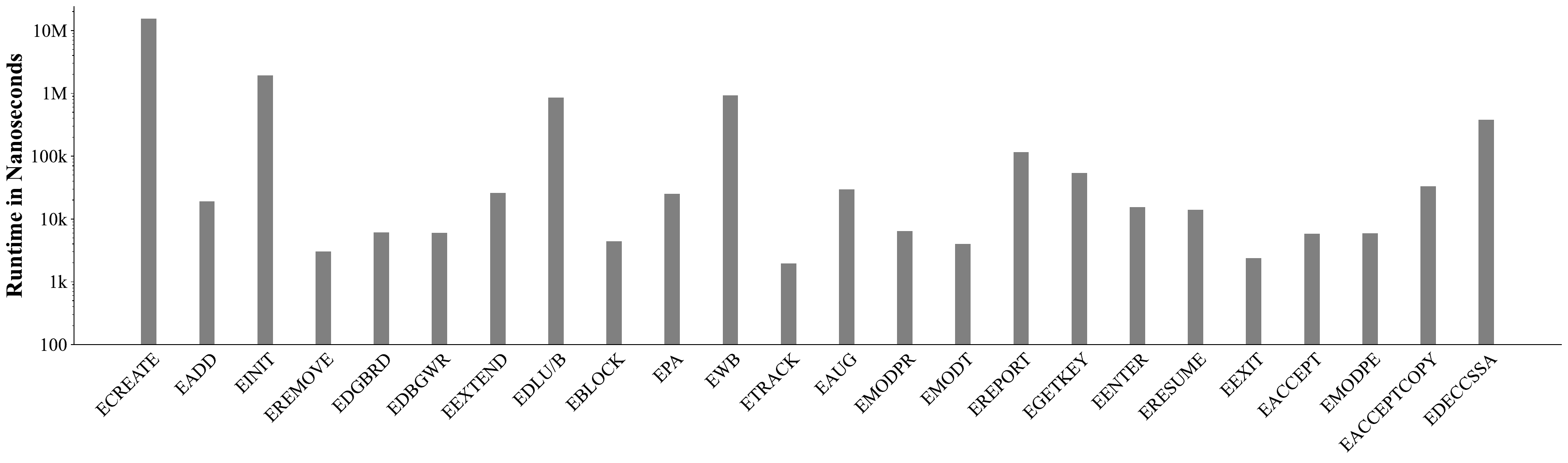}
  \end{center}
  \caption{Results of the microbenchmarks for the enclave microprograms. 
  The y-axis shows the runtime of each microprogram for CCX in nanoseconds on a logarithmic scale. }
  \label{fig:encl_eval}
\end{figure*}

While emulators can be used to demonstrate the functional correctness of CCX, using them to evaluate its performance is not viable, as both QEMU and FVPs are not cycle accurate~\cite{DBLP:conf/uss/LiljestrandNWPE19,DBLP:conf/uss/ZhangHNZLHYH23,DBLP:conf/uss/SridharaBSKAS24}, reducing the meaningfulness of any results. 
To evaluate CCX's performance, we hence implemented it on a Nitrogen8M development board, running an NXP i.MX8M Armv8-A CPU (not supporting CCA) based on a quad-core Arm Cortex-A53 running at 1.5 GHz and 2GB of RAM.
Here, the prototype is built on TF-A version 2.2 and Linux 6.1.22.
The typical approach used in recent work is to implement dummy functions that mimic the functionality of the firmware functions for the novel hardware features~\cite{DBLP:conf/uss/ZhangHNZLHYH23}. 
As this has proven to give the most accurate results, we also follow this approach. 
To this end, we reimplement the exact GPT structure in the board's TF-A, which emulates the assignment and unassignment of GPT pages. 
Accesses to GPT registers are replaced with accesses to debug registers to emulate the access times. 
With this, we evaluate CCX's performance using microbenchmarks for the \texttt{ENCLS} and \texttt{ENCLU} microprograms, and run NBench~\cite{nbench} to assess broader performance.

\textbf{Microbenchmarks.} 
For the microbenchmarks, we execute each microprogram 100 times in CCX and calculate the average execution times.
We note that the variance is omitted since it was negligible across runs.
To accurately measure the timings, we use the Linux kernel's \texttt{ktime} API for \texttt{ENCLS} calls, and the standard library's clock interface (\texttt{clock\_gettime}) with the real-time clock for \texttt{ENCLU} instructions. 

Comparing the results of our microbenchmarks with those of related approaches (e.g., \vSGX~\cite{DBLP:conf/sp/ZhaoLZL22} or \NestedSGX~\cite{DBLP:conf/ndss/0001SMLZY0M025}) or native SGX is challenging.
For example, \NestedSGX could directly compare its performance to \vSGX by running both approaches on the same CPU, but since CCX is the first approach bringing SGX to Arm, no comparable design exists. 
Additionally, scaling the results across different CPUs is not a sound approach, as it would require accounting for the effects of various microarchitectural components (e.g., caches, TLBs, and branch predictors), which is infeasible.
For these reasons, we avoid direct quantitative comparisons of the microbenchmark results.

Figure~\ref{fig:encl_eval} presents the results of our microbenchmarks, while Table~\ref{tab:microinstr} in the appendix reports the absolute values. 
As shown in the figure, some CCX instructions are computationally more expensive than others.
Notably, the \texttt{ECREATE} function is an obvious outlier, requiring over 15~ms to execute. 
This is primarily caused by creating a new GPT for the enclave used in the multi-GPT setup. 
Note that CCX naively sets up new GPTs from scratch, while related work has shown that with so-called \emph{shadow GPTs}, the creation time can be substantially reduced~\cite{DBLP:conf/uss/ZhangHNZLHYH23}. 
However, as this would require additional logic, CCX intentionally sets up GPTs from scratch to keep the TCB minimal.

Instructions related to cryptographic functions (e.g., page encryption and decryption, signature verification, and key derivation) are expensive, as in SGX.
Cryptography is primarily used in two situations: enclave startup (\texttt{EADD}, \texttt{EEXTEND}, and \texttt{ECREATE}) and page swapping (\texttt{EWB} and \texttt{ELDB/U}). 
However, when using CCX, the need for the \texttt{EWB} and \texttt{ELDB/U} functions, and thereby, this bottleneck is removed, which has been cited for causing major performance issues on memory-expensive tasks~\cite{DBLP:journals/csur/WillM23}.
Additionally, since enclave startup only happens once, we believe this overhead is negligible in practice.

The round-trip time of entering and exiting an enclave can be calculated via the values of \texttt{EENTER} (15,153~ns) and \texttt{EEXIT} (2,334~ns), amounting to approximately 17.5~µs.
Since enclaves must be exited and re-entered for system calls or to interact with their U-App, this overhead applies to any such interaction.
However, we argue that the 17.5~µs overhead is minimal and does not significantly degrade performance for most workloads.

\begin{figure*}
  \begin{center}
    \includegraphics[width=0.9\linewidth]{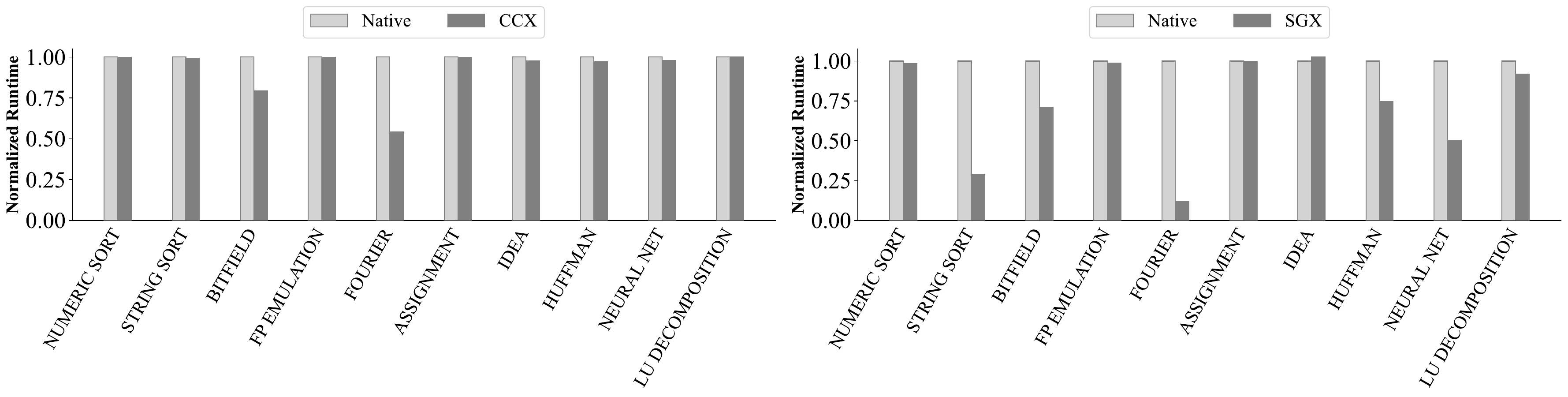}
  \end{center}
  \caption{Results of NBench. Each group on the x-axis corresponds to a benchmark, showing both the native and enclave versions. 
  The y-axis shows the performance of each variant normalized to the native performance.
  The left graph shows the comparison for CCX on Arm hardware, while the right shows the results for SGX on Intel hardware.}
  \label{fig:nbench_eval}
\end{figure*}

\textbf{NBench Benchmarking Tool.} 
To evaluate the performance implications of CCX on a real-world application, we evaluate it using NBench~\cite{nbench} and an SGX-port of it~\cite{sgx-nbench}, as these benchmarks are widely used in related work on SGX enclaves~\cite{DBLP:conf/raid/FuBQL17,DBLP:conf/eurosys/ZhaoLQQ20,DBLP:conf/sp/ZhaoLZL22,DBLP:conf/ndss/0001SMLZY0M025,DBLP:conf/usenix/JiaLWCZYH22}.
We first run the native NBench on the development board to establish a baseline. 
Then, we execute the SGX-port of NBench and compare the results, as shown on the left side of Figure~\ref{fig:nbench_eval}.
For reference, we also run both NBench and its SGX port on an Intel server with an Intel Xeon Silver 4509Y CPU and 128 GB of RAM.
The results are shown on the right side of Figure~\ref{fig:nbench_eval}.

Overall, CCX achieves near-native performance, with two notable exceptions: the \texttt{BITFIELD} and \texttt{FOURIER} test cases show slower results under CCX. 
This is likely due to the custom standard library included in the SGX SDK, which sometimes uses different implementations than glibc. 
This means we experience performance gains when the SGX SDK includes more optimized implementations and performance losses when it lacks them. 
These two test cases rely heavily on certain mathematical operations (e.g., \texttt{cos} and \texttt{sin}), where our implementations in the trusted standard library are apparently slower.
Such performance drops could be addressed by optimizing these functions in future versions. 

On Intel hardware, we observe a significantly larger performance overhead, particularly for memory-intensive workloads such as \texttt{STRING SORT}.
Our results are consistent with prior work evaluating \tool{sgx-nbench}~\cite{DBLP:conf/ndss/0001SMLZY0M025}.
Generally, CCX's performance is, in most cases, better or approximately equal to the performance of SGX. 

\textbf{Comparison with Related Work.}
When excluding CCX benchmarks with high overhead from unoptimized standard library implementations, CCX enclaves incur approximately 1\% overhead. 
In contrast, \vSGX reports an average overhead of over 200\% for NBench~\cite{DBLP:conf/sp/ZhaoLZL22}, while \HyperEnclave and \NestedSGX show similar performance to CCX~\cite{DBLP:conf/usenix/JiaLWCZYH22,DBLP:conf/ndss/0001SMLZY0M025}. 

Regarding Arm enclaves, \Shelter reports a performance overhead of 5.2--15\% across different workloads~\cite{DBLP:conf/uss/ZhangHNZLHYH23}, while NanoZone~\cite{DBLP:journals/corr/abs-2506-07034} reports a 20\% overhead. 
The lowest overhead of 3\% is reported by HiveTEE~\cite{huang2026hivetee}.

In summary, these results show that CCX's performance is better or equal to existing SGX reimplementations and other enclave designs for intra-process isolation, making its application viable in a real-world setting.

\subsection{Goal~3: TCB Increase}

The core idea of CCX is to move the microprograms of Intel SGX into firmware. 
This requires adding a substantial amount of code to the firmware, which runs at the highest privilege level, thereby enlarging the system's TCB. 
To evaluate the dimensions of these additions, we evaluate the SLOC added to the Monitor using cloc 2.0\texttt{cloc}.
Table~\ref{tab:sloc} summarizes how many SLOC were added by the different components.

\begin{table}
  \caption{SLOC of the different CCX components.}
    \centering
    \resizebox{0.8\columnwidth}{!}{
\begin{tabularx}{\columnwidth}{Xr}
  \hline 
  \Tstrut
  \textbf{Component}            &  \textbf{SLOC} \Bstrut \\ \hline \Tstrut
  \textbf{CCX-Core}             & 2,341                \\
  \textbf{CCA-Backend}          & 337  \\
  \textbf{CPUID-Handler}        & 120 \\
  \textbf{SMC/Trap-Handler}     & 149 \\
  \textbf{EPC Management}       & 28 \\
  \textbf{EPCM Management}      & 56 \\
  \textbf{Crypto Management}    & 48 \\
  \textbf{Multi GPT Management} & 106 \\
  \textbf{Utils}                & 366 \\ 
  \textbf{Headers}              & 911 \Bstrut \\ \hline \Tstrut
  \textbf{Total}                & 4,462 \Bstrut \\ \hline 
\end{tabularx}}
\label{tab:sloc}
\end{table}

\begin{table}
  \caption{TCB of related approaches.}
    \centering
    \resizebox{0.8\columnwidth}{!}{
\begin{tabularx}{\columnwidth}{Xr}
  \hline 
  \Tstrut
  \textbf{Approach}                                       &  \textbf{SLOC} \Bstrut \\ \hline \Tstrut
  \textbf{CCX}                                            & 4,462  \\
  \textbf{NestedSGX}~\cite{DBLP:conf/ndss/0001SMLZY0M025} & 5,500  \\
  \textbf{vSGX}~\cite{DBLP:conf/sp/ZhaoLZL22}                       & 9,090  \\
  \textbf{HyperEnclave}~\cite{DBLP:conf/usenix/JiaLWCZYH22}        & 13,719 \Bstrut \\ \hline
\end{tabularx}}
\label{tab:tcbs}
\end{table}

Because we reuse the mbedtls library already shipped with our version of TF-A, its SLOC is not counted toward CCX's TCB. 
Building on the existing Monitor also allows us to leverage many features, such as context management, without increasing the TCB.
In total, we add 4,462 SLOC for CCX. 
The TF-A implementation we used contains about 593k SLOC, of which approximately 143k SLOC are part of mbedtls. 
Therefore, with CCX, we increase the TCB by less than 1\%. 
Table~\ref{tab:tcbs} compares this to related approaches. 
Since the \HyperEnclave approach did not report its TCB in the original publication~\cite{DBLP:conf/usenix/JiaLWCZYH22}, we manually evaluated the TCB using \texttt{cloc} and the master branch of the \HyperEnclave GitHub repository~\cite{hyperenclave_git}, only considering the \emph{./src} subfolder.
Among SGX reimplementations, CCX introduces the smallest increase in TCB.
This is especially significant since CCX implements the full SGX architecture, including the AEX-Notify extension.
On the other hand, e.g., \NestedSGX, which is the closest in TCB increase, does not implement this extension.
However, when the full TCB is considered, the results change.
For CCX, as mentioned above, the entire TF-A must be considered as the TCB.
Nonetheless, related approaches face similar challenges:
\NestedSGX extends the Linux Secure VM Service Modul~\cite{linux_svm} and \vSGX builds on a Linux kernel.
Because both tools run enclaves in AMD SEV VMs, they must also trust the AMD SP co-processor, which contains a separate OS~\cite{DBLP:conf/ccs/BuhrenWS19}.
In this sense, CCX arguably defines a smaller and clearer security boundary.

Only \HyperEnclave differs, as it contains a lightweight hypervisor that uses virtualization to provide enclave functionality. 
However, this approach is least directly comparable because it does not use hardware-based TEEs for the enclaves and therefore provides different security guarantees.

\section{Security Analysis}
\label{sec:sec_discussion}

The third goal of CCX is to provide security guarantees comparable to those of Intel SGX. 
To assess this, we analyze how CCX addresses known classes of SGX attacks, using the taxonomy and survey by Nilsson et al.~\cite{DBLP:journals/corr/abs-2006-13598} as a reference.
We exclude hardware attacks, as they are out of scope.

\subsection{EPC Memory Safety}
\noindent \textbf{Protection against Privileged System Components.}
While SGX guarantees that no untrusted software can access this information via the EPC, CCX achieves this isolation using the GPT. 
However, unlike SGX, where no other system software is trusted, CCX's Monitor at EL3 is trusted and can access enclave memory, increasing CCX's TCB. 
Normally, the RMM can also access realm (and thus enclave) memory, but CCX mitigates this risk through its multi-GPT approach, marking enclave pages as inaccessible in the GPT used by the RMM. 
Thus, only EL3 has access to enclave data, which is unavoidable because it controls the GPT. 

While both approaches rely on a privileged, trusted component (microcode in SGX and EL3 firmware in CCX), the attack surfaces differ in practice. 
While some attacks against microcode have been demonstrated in the past~\cite{DBLP:conf/uss/KoppeKFKGPH17,DBLP:conf/sp/BorrelloESC023}, such attacks are generally rare.
On the other hand, firmware-level components such as EL3 are generally larger and more frequently targeted than CPU microcode, evidenced by the number of vulnerabilities found in Monitor implementations~\cite{lindenmeier2024el3xir}.
Therefore, this may result in a higher practical risk despite similar conceptual trust assumptions.
Nonetheless, since EL3 is, by design, the root of trust on Arm-based systems, it must be trusted when using the GPT. 
With this, CCX trusts only the minimal set of components required on Arm platforms.

\textbf{Memory Encryption.}
Intel SGX employs the \emph{Memory Encryption Engine} (MEE) to ensure EPC integrity and confidentiality~\cite{DBLP:journals/iacr/CostanD16}.
Similarly, Arm CCA mandates memory encryption for realm, root, and SW memory with \texttt{FEAT\_RME}~\cite{arm_mpe}, implemented by the so-called \emph{Memory Protection Engine} (MPE).
In addition to confidentiality, the MPE may perform integrity checks (correctness and freshness) on reads from encrypted memory.
By default, the MPE uses a single per-boot encryption key for all realm memory, but \texttt{FEAT\_MEC} supports individual encryption contexts and keys for different memory regions. 
Although \texttt{FEAT\_MEC} is not yet supported by available hardware or QEMU, CCX can be extended to use it once supported, enabling per-enclave memory encryption keys. 
Therefore, this is only a temporary limitation. 
Additionally, the memory encryption in \texttt{FEAT\_RME} is sufficient to enable protection guarantees similar to SGX.

CCA's threat model explicitly considers some physical memory attacks, such as memory probing and DMA attacks, as well as side-channels like Rowhammer-style attacks~\cite{arm_cca_sec_model}.
While the protection against such attacks is implementation-defined, CCA defines clear guidelines for the guarantees required. 
However, there is no information about exact implementations due to the lack of real-world CCA hardware. 
Nonetheless, as SGX's threat model is also rather ambiguous in this regard~\cite{DBLP:journals/iacr/CostanD16}, no direct comparison is possible.

\textbf{TOCTOU Attacks}
Both SGX and CCX prevent TOCTOU attacks against the microprograms by enforcing a strict ordering in their access semantics.
Any critical information used by the microinstructions is first protected against untrusted access before processing.
SGX achieves this by copying pages to its EPC, whereas CCX protects the underlying memory regions with the GPT.
With this, CCX mitigates TOCTOU attacks within the scope of its microprogram execution.

\textbf{Cross-Enclave Protection.}
SGX uses the \emph{Page Miss Handler} to prevent unauthorized access to EPC pages, ensuring that only the owning enclave can access them.
In CCX, we use a multi-GPT approach~\cite{DBLP:conf/uss/ZhangHNZLHYH23} where an enclave's pages are marked as accessible only in the GPT of the core currently executing the enclave.
Any attempt to access another enclave's memory thus results in a GPF.

\subsection{Control-Flow Attacks}
A major attack targeting Intel SGX is SGX-Step by van Bulck et al.~\cite{DBLP:conf/sosp/BulckPS17}. 
Here, interrupts are configured so that untrusted software can interrupt the enclave after each instruction, effectively enabling single-stepping. 
To mitigate this threat, CCX relies on the AEX-Notify feature~\cite{DBLP:conf/uss/ConstableBCXXAK23}, introduced for SGX, which mitigates such control-flow attacks without impacting overall system functionality.
However, if this feature were to be proven insufficient, the firmware-based implementation of CCX also offers the possibility to implement other defenses:
Unlike SGX, CCX’s firmware-level implementation enables centralized control over interrupt delivery.
With this, CCX can also either completely disable problematic interrupts or add a randomized delay before forwarding them to the NW OS for handling.
However, since such an approach would alter the system's expected behavior, these defenses would have to be carefully designed and are left for future research.

\subsection{Cache Attacks}
Intel SGX has been targeted by a wide range of cache attacks~\cite{DBLP:conf/eurosec/GotzfriedESM17,DBLP:conf/woot/BrasserMDKCS17,DBLP:conf/ches/MoghimiIE17}. 
Since these are enabled by the platform's cache architecture, cache attacks can only be fully prevented by using a different hardware architecture or by disabling caches for enclaves~\cite{DBLP:conf/woot/BrasserMDKCS17}, which is usually ruled out due to the resulting performance overhead. 

CCX does not inherently mitigate cache-based side channels and inherits the platform's underlying cache architecture, similar to SGX. 
While CCA hardware may introduce mitigations, their effectiveness remains to be validated once actual CCA hardware becomes available.

\subsection{Branch Prediction and Speculative Execution Attacks}
Intel CPUs do not clear the branch target buffer on context switches to or from enclaves~\cite{DBLP:conf/uss/0001SGKKP17}, enabling attackers to infer fine-grained enclave control flow.
Similarly, the branch predictor is also not cleared on the context switch, which BranchScope~\cite{DBLP:conf/asplos/EvtyushkinRAP18} and BlueThunder~\cite{DBLP:journals/tches/HuoMWHZZL20} exploit. 
Since Arm CPUs also contain similar branch predictors, this poses a threat to CCX enclaves.
Furthermore, since Spectre~\cite{DBLP:conf/sp/KocherHFGGHHLM019} and Meltdown~\cite{DBLP:conf/uss/Lipp0G0HFHMKGYH18}, research has shown that speculative execution can be used to attack SGX enclaves~\cite{DBLP:journals/ieeesp/ChenCXZLL20}.
Arm CPUs have also been vulnerable to Spectre-like attacks~\cite{DBLP:conf/uss/CanellaB0LBOPEG19,DBLP:conf/dimva/HetterichS22}, again introducing a threat against CCX enclaves.

CCX attempts to mitigate these threats in different ways.
First, CCA hardware aims to mitigate such attacks, making the corresponding hardware features mandatory for its platforms (\texttt{FEAT\_CSV3}). 
Furthermore, CCX uses another upcoming mandatory Arm feature, \texttt{FEAT\_SB}, to add \emph{synchronization barriers} at enclave entry and exit.
However, it must be noted that it's nearly impossible to make absolute statements about protection against such side-channel attacks, since most modern hardware countermeasures have proven insufficient over time, posing a threat to both Intel and Arm platforms.

Nonetheless, if these hardware-based measures are proven insufficient, CCX enables other options:
Its firmware-based design enables highly flexible software-based protections against such attacks~\cite{DBLP:conf/asiaccs/Hetterich00R24}.
This flexibility represents a key advantage over SGX’s microcode-based design, where deploying new defenses is significantly more constrained.

\section{Discussion and Limitations}  
\label{sec:discussion}

In this section, we reflect on the performance and security of CCX and discuss the potential limitations of our prototype.

\textbf{Invasiveness.} 
Since CCX uses the multi-GPT architecture for enclave isolation, it does not interfere with any other existing firmware feature. 
Most importantly, realm VMs can run in parallel with CCX enclaves. 
In such a scenario, if a cVM is scheduled, the multi-GPT architecture enables the \emph{system GPT}, which is used by the RMM and grants it full control over the cVMs.
With this, CCX is completely compatible with the existing CCA architecture. 
Due to this non-invasiveness, implementing CCX on a system is no more invasive than adding any other new functionality to the firmware.

However, a minor limitation arises from the \texttt{ENCLU} gadget, which requires disabling EL0 access to certain features. 
While this design still introduces some transparency issues, we believe the impact can be minimized by choosing features that are rarely used at EL0. 
PMU registers are rarely accessed directly from user space in standard workloads, making them a practical candidate for reuse in our design.

\textbf{Lack of CCA-enabled Hardware.} 
As noted throughout the paper, a limitation of CCX is the lack of publicly available CCA-enabled hardware platforms for evaluation. 
Therefore, we instead used the approach of using an emulated functional prototype and a hardware-based performance prototype using replacements for missing hardware features.
However, such an evaluation setup introduces some uncertainty about the real-world attributes of CCX, for example, when it comes to the performance overhead of memory encryption and its actual implications to security. 
Nonetheless, the approach we used for evaluation is used by virtually \emph{all} existing CCA research~\cite{DBLP:conf/eurosp/BertschiS25}.

\textbf{Performance.}
Most CCX benchmarks run at near-native speed, except for cases that rely on less-optimized mathematical functions. 
Replacing these with Arm-optimized versions should further improve performance in these cases. 

\textbf{Security.}
Although SGX's trust boundary appears smaller, CCX provides the minimum possible trust boundary on Arm platforms without introducing invasive changes. 
Given that the security of CCX enclaves depends on the CCX firmware, future efforts could formally verify it.
This could be achieved using bounded model checking, which Arm also uses to prove the security of other firmware components~\cite{DBLP:journals/pacmpl/FoxSXBMPC23,DBLP:conf/sas/WuXMSC24,DBLP:conf/osdi/LiLDGNSS22}.
Furthermore, while Intel's microcode can be updated, we argue that CCX is easier to update because it is completely software-based. 
This makes it easier to apply mitigations for vulnerabilities or implement new features. 
Finally, the addition of CCX to the TF-A also increases the TCB of other system components requiring trust in EL3, such as the SW or cVMs.
While we argue that the relative TCB increase of 1\% is minimal, further verification of CCX would also reduce the risk introduced for these components.

The lack of real-world Arm CCA hardware makes some aspects of our security analysis theoretical. 
While we trust that the hardware will comply with the CCA requirements, we currently have no way to empirically verify this aspect. 

\textbf{Remote Attestation.}
CCX does not yet support remote attestation due to its reliance on Intel's provisioning and verification architecture~\cite{DBLP:journals/iacr/CostanD16}. 
Solutions like \vSGX can circumvent this issue by leveraging the existing AMD SEV remote attestation ecosystem.
However, there is currently no such ecosystem for Arm CCA. 
This lack of an attestation mechanism is also why related work on Arm enclaves, such as \Shelter~\cite{DBLP:conf/uss/ZhangHNZLHYH23}, implements only local attestation, leaving remote attestation as future work. 
However, CCX goes a step further by already offering a path to design a remote attestation architecture for it once CCA hardware becomes publicly available. 

\section{Related Work}
\label{sec:related_work}

\begin{table*}
  \caption{Comparative table of related work. A full circle indicates complete, and a half circle partial support for a feature.}\label{tab:related_work}
  \label{tab:related}
  \centering
  \begin{adjustbox}{width=\textwidth}
  \begin{tabular}{lcccccccr}
    \toprule
    \textbf{Tool} &  \textbf{Arch}   & \textbf{Intra-Process} &\textbf{SGX Support}  & \textbf{Host-Level Enclaves} & \textbf{Non-Invasive} & \textbf{Minimal TCB} &  \textbf{Proctection} & \textbf{Overhead} \\ 
    \midrule
    \Komodo~\cite{DBLP:conf/sosp/FerraiuoloBHP17}        & Arm    & \fullcircle         &   \emptycircle  & \fullcircle  & \emptycircle  & \fullcircle   & TrustZone+Verification & $<$3\%             \\
    \Sanctuary~\cite{DBLP:conf/ndss/BrasserGJSS19}     & Arm     &  \fullcircle      &   \emptycircle  & \fullcircle  & \fullcircle  & \fullcircle  & TrustZone              & --                \\ 
    \Shelter~\cite{DBLP:conf/uss/ZhangHNZLHYH23}       & Arm        & \emptycircle     &   \emptycircle  & \fullcircle  & \fullcircle  & \halfcircle  & GPT                    & 5.2-15\%          \\
    \Tarnhelm~\cite{DBLP:conf/ccs/QuartaIMFGBLVK21}      & Arm        & \emptycircle     &   \emptycircle  & \fullcircle  & \fullcircle  & \emptycircle  & TrustZone              & $>$300\%           \\
    \TrustShadow~\cite{DBLP:conf/mobisys/GuanLXGZYJ17}   & Arm       & \emptycircle      &   \emptycircle  & \fullcircle  & \fullcircle & \halfcircle   & TrustZone              & 8-137\%          \\
    AnyTEE~\cite{DBLP:journals/access/CerdeiraMSP25}         & Arm      &  \fullcircle       &   \halfcircle   & \fullcircle  & \emptycircle  & \emptycircle   & Virtualization         & $<$3\%           \\
    NanoZone~\cite{DBLP:journals/corr/abs-2506-07034}         & Arm    &  \fullcircle       &   \emptycircle   & \fullcircle  & \fullcircle  & \fullcircle   & GPT         & 20\%           \\
    UIEE~\cite{yanuiee}         & Arm       & \emptycircle    &   \emptycircle   & \fullcircle  & \fullcircle  & \fullcircle   & TrustZone         & 2.65\%           \\
    HiveTEE~\cite{huang2026hivetee}         & Arm     &  \fullcircle      &   \emptycircle   & \fullcircle  & \fullcircle  & \fullcircle   & GPT+MTE         & $<$3\%           \\
    FlexClave~\cite{zhou2026flexclave}         & Arm   &  \fullcircle        &   \emptycircle   & \fullcircle  & \fullcircle  & \fullcircle   & GPT          & $<$10\%           \\
    CCAegis~\cite{liu2025more}         & Arm    &  \fullcircle       &   \emptycircle   & \fullcircle  & \fullcircle  & \fullcircle   & GPT         & 0.8-17\%           \\
    \vSGX~\cite{DBLP:conf/sp/ZhaoLZL22}          & AMD x64  &  \fullcircle       &   \fullcircle   & \fullcircle  & \fullcircle  & \emptycircle   & SEV                    & 221\%           \\ 
    \HyperEnclave~\cite{hyperenclave_git}  & x64     &  \fullcircle        &   \fullcircle   & \fullcircle  & \emptycircle  & \fullcircle   & Virtualization         & $<$5\%             \\ 
    \NestedSGX~\cite{DBLP:conf/ndss/0001SMLZY0M025}     & AMD x64   &  \fullcircle      &   \fullcircle   & \emptycircle  & \fullcircle  & \fullcircle  & SEV+VMPLs              & 1.3\%       \\      
    \midrule 
    \textbf{CCX}   & \textbf{Arm}    &   \fullcircle &   \fullcircle   & \fullcircle  & \fullcircle & \fullcircle    & \textbf{GPT}           & \textbf{1\%}     \\
    \bottomrule
  \end{tabular}
  \end{adjustbox}
\end{table*}

The main goal of CCX is to enable the deployment of existing SGX applications on Arm systems without requiring \emph{any} source-code-level modifications, avoiding the expensive person-hours otherwise needed to port the app while not compromising on usability, performance, or security. 
Therefore, we analyzed existing approaches that implement enclaves for Arm and x64 to conduct a gap analysis. 
Table~\ref{tab:related} shows the results.
\emph{Host-Level Enclaves} refers to the ability for the platform's host (i.e., hypervisor or OS) to spawn enclaves. 
We consider a design \emph{Non-Invasive} when its implementation does not require significant changes to the system's architecture. 
Finally, a design has a \emph{Minimal TCB} if it requires trusting only the minimal set of software and firmware needed to provide isolation.

For Arm, several approaches implement enclaves enabling inter-process isolation~\cite{DBLP:conf/uss/ZhangHNZLHYH23,DBLP:conf/ccs/QuartaIMFGBLVK21,DBLP:conf/mobisys/GuanLXGZYJ17,yanuiee}.
However, as described above, the coarse-grained nature of inter-process isolation does not enable the minimal trust boundaries offered by intra-process isolation.

Nonetheless, there are other approaches that implement intra-process isolation for Arm-based systems.
\Komodo~\cite{DBLP:conf/sosp/FerraiuoloBHP17} aims to create enclaves for Armv7-A CPUs using TrustZone and formal verification. 
For this, it requires replacing the existing EL3 Monitor with its minimal \Komodo Monitor. 
Since on Arm the Monitor is usually expected to handle other features (e.g., PSCI, SCMI, CCA), using \Komodo in a real-world setting is highly invasive.
\Sanctuary~\cite{DBLP:conf/ndss/BrasserGJSS19} also creates user-space enclaves, relying on the TZC-400 to restrict access to \Sanctuary resources to specific CPU cores dedicated to \Sanctuary apps.
Since these cores are no longer available to the NW OS, this approach is also invasive.
It also partially disables caches for enclaves, resulting in a significant performance overhead.
Finally, AnyTEE~\cite{DBLP:journals/access/CerdeiraMSP25} uses virtualization to implement software-defined TEEs in the NW. 
AnyTEE also showcases a minimal SGX runtime that is incomplete and omits most complex features. 
Moreover, this requires replacing the hypervisor, making the approach invasive.
Furthermore, using NW virtualization results in a significant TCB, as all firmware of other security states is trusted, and it also misses out on any hardware protection offered by actual TEEs. 

Similar to CCX, the concurrent works of NanoZone~\cite{DBLP:journals/corr/abs-2506-07034}, FlexClave~\cite{zhou2026flexclave}, and CCAegis~\cite{liu2025more} also reuse the GPT to enable fine-grained intra-process isolation.
However, all of them propose custom programming models and development environments.
Since most apps seeking intra-process isolation currently rely on SGX, any such app must be extensively modified to use these approaches, resulting in expensive development costs. 
This is also why using more flexible TEE SDKs (such as Open Enclave~\cite{open_enclave}) is insufficient, as the existing enclave must still be adapted. 
On the other hand, CCX brings SGX's mature design and programming model to Arm CPUs, enabling straightforward migration of SGX apps.

Alongside these Arm-targeting approaches, some x64-targeting approaches actually succeed in running SGX enclaves on other platforms, making them highly related to CCX.
These approaches usually target AMD and rely on virtualization~\cite{DBLP:conf/sp/ZhaoLZL22,DBLP:conf/usenix/JiaLWCZYH22,DBLP:conf/ndss/0001SMLZY0M025}, but this introduces trade-offs:
\vSGX~\cite{DBLP:conf/sp/ZhaoLZL22} creates new SEV cVMs for each enclave, each containing a small Linux kernel, thereby introducing significant performance overhead, violating our performance goal, and making this approach generally unfavorable.
On the other hand, \HyperEnclave~\cite{DBLP:conf/usenix/JiaLWCZYH22} uses a specialized hypervisor to isolate enclaves. 
Similar to AnyTEE, this limits the hardware protection of enclave pages, makes this system invasive, and, on Arm, would also lead to a large TCB.  
Finally, \NestedSGX~\cite{DBLP:conf/ndss/0001SMLZY0M025} uses VMPLs within an SEV cVM to control enclaves, but cannot spawn user-space enclaves in the host, violating our compatibility goal. 
Therefore, porting these approaches to Arm CCA would fail to sufficiently fulfill our goals.
In addition to these research designs, Amazon Web Services offers \emph{Nitro Enclaves}~\cite{aws}, which rely on virtualization via the \emph{Nitro Hypervisor}. 
In addition to the aforementioned issues with virtualization, it is not compatible with SGX, requiring source code changes to port enclaves to it.

In general, reliance on virtualization does not accurately reflect the SGX concept, whereas CCX provides a more accurate SGX implementation built into Arm's firmware. 
While this is also motivated by the inability to use the hypervisor-level for this task on Arm platforms (see Section~\ref{sec:design}), the firmware on Arm systems is more closely related to the x64 microcode than to the hypervisor privilege level. 
This gives CCX complete control over the hardware, allowing maximum flexibility in using the TEE.
Within x64 platforms (i.e., AMD), approaches for porting the toolchain benefit from the existing SGX SDK, kernel driver, and microcode, making it easier to support SGX library OSes like \Gramine~\cite{DBLP:conf/eurosys/TsaiABJJJKKOP14,DBLP:conf/usenix/TsaiPV17} or \Occulum~\cite{DBLP:conf/asplos/ShenTCCWXXY20} with minimal modification.
However, as these library OSes are developed for x64, porting them to CCX requires additional engineering effort and is left as future work.

In addition to these primary-related approaches, other recent papers highlight the versatility of Arm CCA hardware primitives across diverse applications. 
RContainer~\cite{zhou2025rcontainer} leverages CCA to isolate container runtimes, \BarriCCAde~\cite{DBLP:conf/eurosp/SchulzeLR24} to isolate drivers while \Cage~\cite{DBLP:conf/ndss/WangZDL0NYH24} \textsc{ACAI}~\cite{DBLP:conf/uss/SridharaBSKAS24} and \textsc{Portal}~\cite{sang2025portal} use the GPT to isolate accelerators.
Finally, \textsc{HitchHiker}~\cite{DBLP:conf/ccs/ZhangZZAZ0L24} leverages GPT-protected enclave-like environments for secure logging of compromised OSes.

Finally, on RISC-V, \textsc{Sanctum}~\cite{DBLP:conf/uss/CostanLD16} and \textsc{Cure}~\cite{DBLP:conf/uss/BahmaniBDJKSS21} achieve SGX-comparable enclave security by augmenting the RISC-V architecture, while \textsc{Keystone}~\cite{DBLP:conf/eurosys/LeeKSAS20} implements a framework for custom TEE deployment without hardware modification. 

\section{Conclusion}
\label{sec:conclusion}
In this work, we presented CCX, the first framework that brings the mature design of SGX's intra-process isolation to Arm platforms.
By enabling unmodified Intel SGX applications to run on top of it, CCX also enables straightforward porting of the existing ecosystem of SGX applications to modern Arm systems.
CCX transparently emulates SGX microcode in Arm firmware and reroutes all SGX-related instructions at both the application and OS levels to it.
We implemented CCX for the QEMU emulator and a Nitrogen8M development board, showing that real-world SGX applications execute with near-native performance while providing security guarantees comparable to Intel SGX.
In the future, its firmware-based design enables flexible mitigation of potential vulnerabilities and easy extension of CCX with new enclave features.

\balance
\bibliographystyle{plain}
\bibliography{strings,bibliography}

@misc{cloc,
    title = {{cloc}},
    howpublished = {\url{https://github.com/AlDanial/cloc/tree/v2.00}},
    year = {2024},
    note = {Accessed: 2025-08-25}
}

@article{DBLP:journals/csur/PintoS19,
  author       = {Sandro Pinto and
                  Nuno Santos},
  title        = {{Demystifying Arm TrustZone: A Comprehensive Survey}},
  journal      = acm-csur,
volume = {51},
number = {6},
  year         = {2019},
}

@misc{arm_mpe,
    title = {{Arm Realm Management Extension (RME) System Architecture}},
    howpublished = {\url{https://developer.arm.com/documentation/den0129/latest/}},
    year = {2023},
    note = {Accessed: 2025-08-25}
}

@misc{arm_cca_sec_model,
    title = {{Arm CCA Security Model 1.0}},
    howpublished = {\url{https://documentation-service.arm.com/static/610aaec33d73a34b640e333b?token=}},
    year = {2021},
    note = {Accessed: 2025-08-25}
}

@inproceedings{DBLP:conf/osdi/LiLDGNSS22,
  author       = {Xupeng Li and
                  Xuheng Li and
                  Christoffer Dall and
                  Ronghui Gu and
                  Jason Nieh and
                  Yousuf Sait and
                  Gareth Stockwell},
  title        = {{Design and Verification of the Arm Confidential Compute Architecture}},
  booktitle    = osdi,
  year         = {2022},
}

@article{DBLP:journals/pacmpl/FoxSXBMPC23,
  author       = {Anthony C. J. Fox and
                  Gareth Stockwell and
                  Shale Xiong and
                  Hanno Becker and
                  Dominic P. Mulligan and
                  Gustavo Petri and
                  Nathan Chong},
  title        = {{A Verification Methodology for the Arm{\textregistered} Confidential Computing Architecture: From a Secure Specification to Safe Implementations}},
  journal      = oopsla,
  year         = {2023},
}

@inproceedings{DBLP:conf/seed/MulliganPSSV21,
  author       = {Dominic P. Mulligan and
                  Gustavo Petri and
                  Nick Spinale and
                  Gareth Stockwell and
                  Hugo J. M. Vincent},
  title        = {{Confidential Computing - a brave new world}},
  booktitle    = seed,
  year         = {2021},
}

@inproceedings{DBLP:conf/sas/WuXMSC24,
  author       = {Tong Wu and
                  Shale Xiong and
                  Edoardo Manino and
                  Gareth Stockwell and
                  Lucas C. Cordeiro},
  title        = {{Verifying Components of Arm\({}^{\mbox{{\textregistered}}}\) Confidential
                  Computing Architecture with {ESBMC}}},
  booktitle    = sas, 
  year         = {2024},
}

@inproceedings{DBLP:conf/uss/SridharaBSKAS24,
  author       = {Supraja Sridhara and
                  Andrin Bertschi and
                  Benedict Schl{\"{u}}ter and
                  Mark Kuhne and
                  Fabio Aliberti and
                  Shweta Shinde},
  title        = {{ACAI: Protecting Accelerator Execution with Arm Confidential Computing
                  Architecture}},
  booktitle    = usenix-security,
  year         = {2024},
}

@inproceedings{DBLP:conf/uss/ZhangHNZLHYH23,
  author       = {Yiming Zhang and
                  Yuxin Hu and
                  Zhenyu Ning and
                  Fengwei Zhang and
                  Xiapu Luo and
                  Haoyang Huang and
                  Shoumeng Yan and
                  Zhengyu He},
  title        = {{{SHELTER:} Extending Arm {CCA} with Isolation in User Space}},
  booktitle    = usenix-security,
  year         = {2023},
}

@inproceedings{DBLP:conf/ndss/WangZDL0NYH24,
  author       = {Chenxu Wang and
                  Fengwei Zhang and
                  Yunjie Deng and
                  Kevin Leach and
                  Jiannong Cao and
                  Zhenyu Ning and
                  Shoumeng Yan and
                  Zhengyu He},
  title        = {{{CAGE:} Complementing Arm {CCA} with {GPU} Extensions}},
  booktitle    = isoc-ndss,
  year         = {2024},
}

@inproceedings{DBLP:conf/eurosp/SchulzeLR24,
  author       = {Matti Schulze and
                  Christian Lindenmeier and
                  Jonas R{\"{o}}ckl},
  title        = {{BarriCCAde: Isolating Closed-Source Drivers with {ARM} {CCA}}},
  booktitle    = ieee-eurospw, 
  year         = {2024},
}

@inproceedings{zhou2025rcontainer,
  title={{RContainer: A Secure Container Architecture through Extending ARM CCA Hardware Primitives.}},
  author={Zhou, Qihang and Cao, Wenzhuo and Jia, Xiaoqi and Liu, Peng and Zhang, Shengzhi and Chen, Jiayun and Xu, Shaowen and Song, Zhenyu},
  booktitle=ndss,
  year={2025}
}

@inproceedings{sang2025portal,
  title={PORTAL: Fast and Secure Device Access with Arm CCA for Modern Arm Mobile System-on-Chips (SoCs)},
  author={Sang, Fan and Lee, Jaehyuk and Zhang, Xiaokuan and Kim, Taesoo},
  booktitle= ieee-sp,
  year={2025}
}

@inproceedings{DBLP:conf/eurosp/BertschiS25,
  author       = {Andrin Bertschi and
                  Shweta Shinde},
  title        = {{{OPENCCA:} An Open Framework to Enable Arm {CCA} Research}},
  booktitle    = {ieee-eurospw},
  year         = {2025},
}

@article{DBLP:journals/corr/abs-2506-07034,
  author       = {Shiqi Liu and
                  Yongpeng Gao and
                  Mingyang Zhang and
                  Jie Wang},
  title        = {{NanoZone: Scalable, Efficient, and Secure Memory Protection for Arm
                  CCA}},
  journal      = {arXiv},
  year         = {2025},
}

@article{huang2026hivetee,
  title={{HiveTEE: Scalable and Fine-grained Isolated Domains with RME and MTE Co-assisted}},
  author={Huang, Haoyang and Zhangy, Fengwei},
  journal={IEEE Transactions on Information Forensics and Security},
  year={2026},
}

@article{liu2025more,
  title={More Granular, Less Trust: Enforcing Intra-Process Isolation With Arm CCA in an Untrusted Management Environment},
  author={Liu, Shiqi and Jiang, Zhouqi and Wang, Jie and Zhou, Wei and Sun, Kun and Chen, Zhaohui and Xie, Yulai},
  journal={IEEE Transactions on Information Forensics and Security},
  volume={20},
  pages={12507--12522},
  year={2025},
  publisher={IEEE}
}

@article{yanuiee,
  title={{UIEE: Secure and Efficient User-space Isolated Execution Environment for Embedded TEE Systems}},
  author={Yan, Huaiyu and Ling, Zhen and Chen, Xuandong and Shao, Xinhui and Jin, Yier and Li, Haobo and Yang, Ming and Jiang, Ping and Luo, Junzhou},
  booktitle=ndss,
  year={2026}
}

@article{zhou2026flexclave,
  title={{FlexClave: An Extensible and Secure Trusted Execution Environment Framework}},
  author={Zhou, Qihang and Cao, Wenzhuo and Jia, Xiaoqi and Xu, Shaowen and Chen, Jiayun and Jiang, Nan and Zhang, Zhicong and Song, Zhenyu and Du, Haichao and Xie, Yamin and others},
  journal={IEEE Transactions on Computers},
  year={2026},
}

@inproceedings{DBLP:conf/uss/LiljestrandNWPE19,
  author       = {Hans Liljestrand and
                  Thomas Nyman and
                  Kui Wang and
                  Carlos Chinea Perez and
                  Jan{-}Erik Ekberg and
                  N. Asokan},
  title        = {{{PAC} it up: Towards Pointer Integrity using {ARM} Pointer Authentication}},
  booktitle    = usenix-security,
  year         = {2019},
}

@inproceedings{anati2013innovative,
  title={Innovative technology for CPU based attestation and sealing},
  author={Anati, Ittai and Gueron, Shay and Johnson, Simon and Scarlata, Vincent},
  booktitle= hasp,
  year={2013},
}

@inproceedings{DBLP:conf/isca/HoekstraLPPC13,
  author       = {Matthew Hoekstra and
                  Reshma Lal and
                  Pradeep Pappachan and
                  Vinay Phegade and
                  Juan del Cuvillo},
  title        = {{Using innovative instructions to create trustworthy software solutions}},
  booktitle    = hasp,
  year         = {2013},
}

@inproceedings{DBLP:conf/isca/McKeenABRSSS13,
  author       = {Frank McKeen and
                  Ilya Alexandrovich and
                  Alex Berenzon and
                  Carlos V. Rozas and
                  Hisham Shafi and
                  Vedvyas Shanbhogue and
                  Uday R. Savagaonkar},
  title        = {{Innovative instructions and software model for isolated execution}},
  booktitle    = hasp, 
  year         = {2013},
}

@techreport{DBLP:journals/iacr/CostanD16,
  author       = {Victor Costan and
                  Srinivas Devadas},
  title        = {{Intel {SGX} Explained}},
  institution      = {{{IACR} Cryptology ePrint Archive}},
  year         = {2016},
  note = {\url{https://eprint.iacr.org/2016/086}}
}

@misc{sgx_sdk,
    title = {{Linux-SGX}},
    howpublished = {\url{https://github.com/intel/linux-sgx}},
    note = {Accessed: 2025-08-25}
}

@article{DBLP:journals/csur/ChengOVAGJFB24,
  author       = {Pau{-}Chen Cheng and
                  Wojciech Ozga and
                  Enriquillo Valdez and
                  Salman Ahmed and
                  Zhongshu Gu and
                  Hani Jamjoom and
                  Hubertus Franke and
                  James Bottomley},
  title        = {{Intel {TDX} Demystified: {A} Top-Down Approach}},
  journal      = acm-csur,
  year         = {2024},
volume = {56},
number = {9},
}

@article{sev2020strengthening,
  title={{Strengthening VM isolation with integrity protection and more}},
  author={{AMD}},
  journal={White Paper},
  year={2020}
}

@inproceedings{DBLP:conf/uss/CostanLD16,
  author       = {Victor Costan and
                  Ilia A. Lebedev and
                  Srinivas Devadas},
  title        = {{Sanctum: Minimal Hardware Extensions for Strong Software Isolation}},
  booktitle    = usenix-security,
  year         = {2016},
}

@inproceedings{DBLP:conf/eurosys/LeeKSAS20,
  author       = {Dayeol Lee and
                  David Kohlbrenner and
                  Shweta Shinde and
                  Krste Asanovic and
                  Dawn Song},
  title        = {{Keystone: an open framework for architecting trusted execution environments}},
  booktitle    = eurosys,
  year         = {2020},
}

@inproceedings{DBLP:conf/uss/BahmaniBDJKSS21,
  author       = {Raad Bahmani and
                  Ferdinand Brasser and
                  Ghada Dessouky and
                  Patrick Jauernig and
                  Matthias Klimmek and
                  Ahmad{-}Reza Sadeghi and
                  Emmanuel Stapf},
  title        = {{{CURE:} {A} Security Architecture with CUstomizable and Resilient
                  Enclaves}},
  booktitle    = usenix-security,
  year         = {2021},
}

@inproceedings{DBLP:conf/ndss/BrasserGJSS19,
  author       = {Ferdinand Brasser and
                  David Gens and
                  Patrick Jauernig and
                  Ahmad{-}Reza Sadeghi and
                  Emmanuel Stapf},
  title        = {{{SANCTUARY:} ARMing TrustZone with User-space Enclaves}},
  booktitle    =  isoc-ndss,
  year         = {2019},
}

@inproceedings{DBLP:conf/sosp/FerraiuoloBHP17,
  author       = {Andrew Ferraiuolo and
                  Andrew Baumann and
                  Chris Hawblitzel and
                  Bryan Parno},
  title        = {{Komodo: Using verification to disentangle secure-enclave hardware
                  from software}},
  booktitle    = sosp,
  year         = {2017},
}

@inproceedings{DBLP:conf/usenix/JiaLWCZYH22,
  author       = {Yuekai Jia and
                  Shuang Liu and
                  Wenhao Wang and
                  Yu Chen and
                  Zhengde Zhai and
                  Shoumeng Yan and
                  Zhengyu He},
  title        = {{HyperEnclave: An Open and Cross-platform Trusted Execution Environment}},
  booktitle    = usenix-atc,
  year         = {2022},
}

@inproceedings{DBLP:conf/sp/ZhaoLZL22,
  author       = {Shixuan Zhao and
                  Mengyuan Li and
                  Yinqian Zhang and
                  Zhiqiang Lin},
  title        = {{vSGX: Virtualizing {SGX} Enclaves on {AMD} {SEV}}},
  booktitle    = ieee-sp, 
  year         = {2022},
}

@inproceedings{DBLP:conf/ndss/0001SMLZY0M025,
  author       = {Wenhao Wang and
                  Linke Song and
                  Benshan Mei and
                  Shuang Liu and
                  Shijun Zhao and
                  Shoumeng Yan and
                  XiaoFeng Wang and
                  Dan Meng and
                  Rui Hou},
  title        = {{The Road to Trust: Building Enclaves within Confidential VMs}},
  booktitle    = isoc-ndss,
  year         = {2025},
}

@inproceedings{DBLP:conf/mobisys/GuanLXGZYJ17,
  author       = {Le Guan and
                  Peng Liu and
                  Xinyu Xing and
                  Xinyang Ge and
                  Shengzhi Zhang and
                  Meng Yu and
                  Trent Jaeger},
  title        = {{TrustShadow: Secure Execution of Unmodified Applications with {ARM}
                  TrustZone}},
  booktitle    = mobisys,
  year         = {2017},
}

@inproceedings{DBLP:conf/ccs/ZhangZZAZ0L24,
  author       = {Chuqi Zhang and
                  Jun Zeng and
                  Yiming Zhang and
                  Adil Ahmad and
                  Fengwei Zhang and
                  Hai Jin and
                  Zhenkai Liang},
  title        = {{The HitchHiker's Guide to High-Assurance System Observability Protection
                  with Efficient Permission Switches}},
  booktitle    = acm-ccs,
  year         = {2024},
}

@article{DBLP:journals/access/CerdeiraMSP25,
  author       = {David Cerdeira and
                  Jos{\'{e}} Martins and
                  Nuno Santos and
                  Sandro Pinto},
  title        = {{AnyTEE: An Open and Interoperable Software Defined {TEE} Framework}},
  journal      = {{IEEE} Access},
  year         = {2025},
}

@inproceedings{DBLP:conf/sp/BorrelloESC023,
  author       = {Pietro Borrello and
                  Catherine Easdon and
                  Martin Schwarzl and
                  Roland Czerny and
                  Michael Schwarz},
  title        = {{CustomProcessingUnit: Reverse Engineering and Customization of Intel
                  Microcode}},
  booktitle    = spw,
  year         = {2023},
}

@inproceedings{DBLP:conf/uss/KoppeKFKGPH17,
  author       = {Philipp Koppe and
                  Benjamin Kollenda and
                  Marc Fyrbiak and
                  Christian Kison and
                  Robert Gawlik and
                  Christof Paar and
                  Thorsten Holz},
  title        = {{Reverse Engineering x86 Processor Microcode}},
  booktitle    = usenix-security,
  year         = {2017},
}

@inproceedings{lindenmeier2024el3xir,
  author       = {Christian Lindenmeier and
                  Mathias Payer and
                  Marcel Busch},
  title        = {{EL3XIR: Fuzzing COTS Secure Monitors}},
  booktitle    = usenix-security,
  year         = {2024},
}

@techreport{DBLP:journals/corr/abs-2006-13598,
  author       = {Alexander Nilsson and
                  Pegah Nikbakht Bideh and
                  Joakim Brorsson},
  title        = {{A Survey of Published Attacks on Intel {SGX}}},
  institution  = {arXiv},
  year         = {2020},
  note={\url{https://arxiv.org/abs/2006.13598}}, 
}

@inproceedings{DBLP:conf/eurosec/GotzfriedESM17,
  author       = {Johannes G{\"{o}}tzfried and
                  Moritz Eckert and
                  Sebastian Schinzel and
                  Tilo M{\"{u}}ller},
  title        = {{Cache Attacks on Intel {SGX}}},
  booktitle    = eurosec,
  year         = {2017},
}

@inproceedings{DBLP:conf/woot/BrasserMDKCS17,
  author       = {Ferdinand Brasser and
                  Urs M{\"{u}}ller and
                  Alexandra Dmitrienko and
                  Kari Kostiainen and
                  Srdjan Capkun and
                  Ahmad{-}Reza Sadeghi},
  title        = {{Software Grand Exposure: {SGX} Cache Attacks Are Practical}},
  booktitle    = usenix-woot, 
  year         = {2017},
}

@inproceedings{DBLP:conf/ches/MoghimiIE17,
  author       = {Ahmad Moghimi and
                  Gorka Irazoqui and
                  Thomas Eisenbarth},
  title        = {{CacheZoom: How {SGX} Amplifies the Power of Cache Attacks}},
  booktitle    = ches,
  year         = {2017},
}

@inproceedings{DBLP:conf/sosp/BulckPS17,
  author       = {Jo Van Bulck and
                  Frank Piessens and
                  Raoul Strackx},
  title        = {{SGX-Step: {A} Practical Attack Framework for Precise Enclave Execution
                  Control}},
  booktitle    = systex,
  year = {2017}
}

@inproceedings{DBLP:conf/uss/0001SGKKP17,
  author       = {Sangho Lee and
                  Ming{-}Wei Shih and
                  Prasun Gera and
                  Taesoo Kim and
                  Hyesoon Kim and
                  Marcus Peinado},
  title        = {{Inferring Fine-grained Control Flow Inside {SGX} Enclaves with Branch
                  Shadowing}},
  booktitle    = usenix-security,
  year         = {2017},
}

@inproceedings{DBLP:conf/asplos/EvtyushkinRAP18,
  author       = {Dmitry Evtyushkin and
                  Ryan Riley and
                  Nael B. Abu{-}Ghazaleh and
                  Dmitry Ponomarev},
  title        = {{BranchScope: {A} New Side-Channel Attack on Directional Branch Predictor}},
  booktitle    = asplos, 
  year         = {2018},
}

@article{DBLP:journals/tches/HuoMWHZZL20,
  author       = {Tianlin Huo and
                  Xiaoni Meng and
                  Wenhao Wang and
                  Chunliang Hao and
                  Pei Zhao and
                  Jian Zhai and
                  Mingshu Li},
  title        = {{Bluethunder: {A} 2-level Directional Predictor Based Side-Channel
                  Attack against {SGX}}},
  journal      = {{IACR Transactions on Cryptographic Hardware and Embedded Systems}},
  volume = {2020},
  number = {1},
  year         = {2020},
}

@inproceedings{DBLP:conf/sp/KocherHFGGHHLM019,
  author       = {Paul Kocher and
                  Jann Horn and
                  Anders Fogh and
                  Daniel Genkin and
                  Daniel Gruss and
                  Werner Haas and
                  Mike Hamburg and
                  Moritz Lipp and
                  Stefan Mangard and
                  Thomas Prescher and
                  Michael Schwarz and
                  Yuval Yarom},
  title        = {{Spectre Attacks: Exploiting Speculative Execution}},
  booktitle    = ieee-sp,
  year         = {2019},
}

@inproceedings{DBLP:conf/uss/Lipp0G0HFHMKGYH18,
  author       = {Moritz Lipp and
                  Michael Schwarz and
                  Daniel Gruss and
                  Thomas Prescher and
                  Werner Haas and
                  Anders Fogh and
                  Jann Horn and
                  Stefan Mangard and
                  Paul Kocher and
                  Daniel Genkin and
                  Yuval Yarom and
                  Mike Hamburg},
  title        = {{Meltdown: Reading Kernel Memory from User Space}},
  booktitle    = usenix-security,
  year         = {2018},
}

@article{DBLP:journals/ieeesp/ChenCXZLL20,
  author       = {Guoxing Chen and
                  Sanchuan Chen and
                  Yuan Xiao and
                  Yinqian Zhang and
                  Zhiqiang Lin and
                  Ten{-}Hwang Lai},
  title        = {{SgxPectre: Stealing Intel Secrets From {SGX} Enclaves via Speculative
                  Execution}},
  journal      = ieee-sp,
  year         = {2020},
}

@inproceedings{DBLP:conf/uss/CanellaB0LBOPEG19,
  author       = {Claudio Canella and
                  Jo Van Bulck and
                  Michael Schwarz and
                  Moritz Lipp and
                  Benjamin von Berg and
                  Philipp Ortner and
                  Frank Piessens and
                  Dmitry Evtyushkin and
                  Daniel Gruss},
  title        = {{A Systematic Evaluation of Transient Execution Attacks and Defenses}},
  booktitle    = usenix-security,
  year         = {2019},
}

@inproceedings{DBLP:conf/asiaccs/Hetterich00R24,
  author       = {Lorenz Hetterich and
                  Markus Bauer and
                  Michael Schwarz and
                  Christian Rossow},
  title        = {{Switchpoline: {A} Software Mitigation for Spectre-BTB and Spectre-BHB
                  on ARMv8}},
  booktitle    = asia-ccs,
  year         = {2024},
}

@inproceedings{DBLP:conf/dimva/HetterichS22,
  author       = {Lorenz Hetterich and
                  Michael Schwarz},
  title        = {{Branch Different - Spectre Attacks on Apple Silicon}},
  booktitle    = dimva,
  year         = {2022},
}

@misc{nbench,
    title = {{NBench}},
    howpublished = {\url{https://www.math.utah.edu/~mayer/linux/bmark.html}},
    year = {2017},
    note = {Accessed: 2025-08-25}
}

@misc{sgx-nbench,
    title = {{ The nbench benchmark ported to SGX. }},
    howpublished = {\url{https://github.com/utds3lab/sgx-nbench}},
    note = {Accessed: 2025-08-25}
}

@misc{mbedtls,
    title = {{ Mbed TLS }},
    howpublished = {\url{https://github.com/Mbed-TLS/mbedtls}},
    note = {Accessed: 2025-08-25}
}

@misc{tfa,
    title = {{ Trusted Firmware-A }},
    howpublished = {\url{https://github.com/ARM-software/arm-trusted-firmware}},
    note = {Accessed: 2025-08-25}
}

@misc{rmm_spec,
    title = {{ Realm Management Monitor
specification }},
    howpublished = {\url{
https://documentation-service.arm.com/static/69cb945ac1586b7c59b1c00c?token=}},
    note = {Accessed: 2026-05-06}
}

@article{DBLP:journals/pomacs/NgocBBTSFH19,
  author       = {Tu Dinh Ngoc and
                  Bao Bui and
                  Stella Bitchebe and
                  Alain Tchana and
                  Valerio Schiavoni and
                  Pascal Felber and
                  Daniel Hagimont},
  title        = {{Everything You Should Know About Intel {SGX} Performance on Virtualized
                  Systems}},
  journal      = acm-pomacs,
  year         = {2019},
}

@misc{sgx_deprecated,
  author = {Intel},
  title={11th Generation Intel{\textregistered} Core™ Processors},
  howpublished = {\url{https://cdrdv2.intel.com/v1/dl/getContent/634648}},
  year={2020},
    note = {Accessed: 2025-08-25}
}

@misc{sgx_prog_model,
    author = {Intel},
    title = {{Intel® Software Guard Extensions (Intel® SGX)}},
    howpublished = {\url{https://cdrdv2-public.intel.com/671581/intel-sgx-developer-guide.pdf}},
    year = {2018},
    note = {Accessed: 2025-08-25}
}

@inproceedings{DBLP:conf/ccs/QuartaIMFGBLVK21,
  author       = {Davide Quarta and
                  Michele Ianni and
                  Aravind Machiry and
                  Yanick Fratantonio and
                  Eric Gustafson and
                  Davide Balzarotti and
                  Martina Lindorfer and
                  Giovanni Vigna and
                  Christopher Kruegel},
  title        = {{Tarnhelm: Isolated, Transparent {\&} Confidential Execution of Arbitrary
                  Code in ARM's TrustZone}},
  booktitle    = {{Research on Offensive and Defensive Techniques in the Context of Man At The End (MATE) Attacks}},
  year         = {2021},
}

@article{DBLP:journals/csur/WillM23,
  author       = {Newton Carlos Will and
                  Carlos Alberto Maziero},
  title        = {{Intel Software Guard Extensions Applications: A Survey}},
  journal      = acm-csur,
  year         = {2023},
volume = {55},
number = {14s},
}

@inproceedings{DBLP:conf/osdi/ConnellFSDP24,
  author       = {Graeme Connell and
                  Vivian Fang and
                  Rolfe Schmidt and
                  Emma Dauterman and
                  Raluca Ada Popa},
  title        = {{Secret Key Recovery in a Global-Scale End-to-End Encryption System}},
  booktitle    = osdi,
  year         = {2024},
}

@misc{signal_contact,
    author = {Signal},
    title = {{Technology preview: Private contact discovery for Signal}},
    howpublished = {\url{https://signal.org/blog/private-contact-discovery/}},
    year = {2017},
    note = {Accessed: 2025-08-25}
}

@misc{arm-marketshare,
author = {{Grand View Research}},
title = {{ARM-Based Servers Market Summary}},
howpublished = {\url{https://www.grandviewresearch.com/industry-analysis/arm-based-servers-market-report}},
year = {2025},
    note = {Accessed: 2025-08-27}
}

@misc{rowhammer,
author ={Mark Seaborn and Thomas Dullien},
title = {Exploiting the {DRAM} rowhammer bug to gain kernel privileges},
howpublished = {\url{https://googleprojectzero.blogspot.com/2015/03/exploiting-dram-rowhammer-bug-to-gain.html}},
year = {2015}
}

@inproceedings{10.5555/3241094.3241096,
author = {Razavi, Kaveh and Gras, Ben and Bosman, Erik and Preneel, Bart and Giuffrida, Cristiano and Bos, Herbert},
title = {{Flip Feng Shui: Hammering a needle in the software stack}},
year = {2016},
booktitle = usenix-security,
}

@inproceedings{10.5555/3241094.3241097,
author = {Xiao, Yuan and Zhang, Xiaokuan and Zhang, Yinqian and Teodorescu, Radu},
title = {{One bit flips, one cloud flops: cross-VM row hammer attacks and privilege escalation}},
year = {2016},
booktitle = usenix-security,
}

@misc{flashbots,
    author = {Flashbots},
    title = {{Block Building inside SGX}},
    howpublished = {\url{https://writings.flashbots.net/block-building-inside-sgx#our-sepolia-sgx-builder}},
    year = {2023},
    note = {Accessed: 2025-08-25}
}

@misc{gramine, 
  author = {Gramine},
  title = {{Users of Gramine}},
  howpublished = {\url{https://gramine.readthedocs.io/en/stable/gramine-users.html}},
  year = {2023},
    note = {Accessed: 2025-04-23}
}

@misc{mozilla, 
  author = {Mozilla},
  title = {{Privacy-preserving digital ads infrastructure: An overview of Anonym’s technology}},
  howpublished = {\url{https://blog.mozilla.org/en/products/anonym-technology-overview/}},
  year = {2025},
    note = {Accessed: 2025-08-25}
}

@misc{cosmian, 
  author = {Cosmian},
  title = {{Technology}},
  howpublished = {\url{https://cosmian.com/technology/}},
  year = {2024},
    note = {Accessed: 2025-08-25}
}

@misc{intel_sgx_tdx, 
  author = {Intel},
  title = {{Securing Your Trust Boundary with Intel SGX and Intel TDX}},
  howpublished = {\url{https://www.intel.com/content/www/us/en/content-details/816053/securing-your-trust-boundary-with-intel-sgx-and-intel-tdx.html?DocID=816053}},
  year = {2024},
    note = {Accessed: 2025-08-25}
}

@misc{ms_payment, 
  author = {Intel},
  title = {{Microsoft Protects \$25B in Customer Payments}},
  howpublished = {\url{https://www.intel.com/content/www/us/en/security/resources/microsoft-azure-confidential-computing-brief.html}},
  year = {2023},
    note = {Accessed: 2025-08-25}
}

@misc{german, 
  author = {Intel},
  title = {{Intel SGX Protects German Electronic Patient Records}},
  howpublished = {\url{https://www.intel.com/content/www/us/en/newsroom/news/intel-sgx-protects-german-electronic-patient-records.html#gs.heuo6x}},
  year = {2021},
    note = {Accessed: 2025-08-25}
}

@inproceedings{DBLP:conf/uss/ConstableBCXXAK23,
  author       = {Scott Constable and
                  Jo Van Bulck and
                  Xiang Cheng and
                  Yuan Xiao and
                  Cedric Xing and
                  Ilya Alexandrovich and
                  Taesoo Kim and
                  Frank Piessens and
                  Mona Vij and
                  Mark Silberstein},
  title        = {{AEX-Notify: Thwarting Precise Single-Stepping Attacks through Interrupt Awareness for Intel {SGX} Enclaves}},
  booktitle = usenix-security,
  year         = {2023},
}

@inproceedings{DBLP:conf/esorics/ChandraKLKKT17,
  author       = {Swarup Chandra and
                  Vishal Karande and
                  Zhiqiang Lin and
                  Latifur Khan and
                  Murat Kantarcioglu and
                  Bhavani Thuraisingham},
  title        = {{Securing Data Analytics on {SGX} with Randomization}},
  booktitle    = esorics,
  year         = {2017},
}

@inproceedings{DBLP:conf/infocom/ZhangLZLWW20,
  author       = {Xiaoli Zhang and
                  Fengting Li and
                  Zeyu Zhang and
                  Qi Li and
                  Cong Wang and
                  Jianping Wu},
  title        = {{Enabling Execution Assurance of Federated Learning at Untrusted Participants}},
  booktitle    = ieee-infocom-mc,
  year         = {2020},
}

@misc{sgx_kmeans, 
  title = {{sgx-kmeans}},
  howpublished = {\url{https://github.com/dsc-sgx/sgx-kmeans}},
  note = {Accessed: 2025-08-25}
}

@misc{sgxsqlite, 
  title = {{SGX\_SQLite}},
  howpublished = {\url{https://github.com/yerzhan7/SGX_SQLite}},
  note = {Accessed: 2025-08-25}
}

@misc{talos, 
  author={{Aublin, Pierre-Louis and Kelbert, Florian and O’keeffe, Dan and Muthukumaran, Divya and Priebe, Christian and Lind, Joshua and Krahn, Robert and Fetzer, Christof and Eyers, David and Pietzuch, Peter}},
  title = {{TaLoS: Efficient TLS Termination Inside SGX Enclaves for Existing Applications}},
  howpublished = {\url{https://github.com/lsds/TaLoS}},
  year = {2017},
  note = {Accessed: 2025-08-25}
}

@misc{nginx, 
  author={{NGINX}},
  title = {{nginx}},
  howpublished = {\url{https://nginx.org/en/}},
  year = {2024},
  note = {Accessed: 2025-08-25}
}

@misc{aws, 
  author={{Amazon Web Services}},
  title = {{The Security Design of the AWS Nitro System: AWS Whitepaper}},
  howpublished = {\url{https://docs.aws.amazon.com/pdfs/whitepapers/latest/security-design-of-aws-nitro-system/security-design-of-aws-nitro-system.pdf}},
  year = {2024},
  note = {Accessed: 2025-08-25}
}

@misc{open_enclave, 
  title = {{Open Enclave SDK}},
  howpublished = {\url{https://github.com/openenclave/openenclave}},
  note = {Accessed: 2025-08-25}
}

@inproceedings{DBLP:conf/raid/FuBQL17,
  author       = {Yangchun Fu and
                  Erick Bauman and
                  Raul Quinonez and
                  Zhiqiang Lin},
  title        = {{SGX-LAPD: Thwarting Controlled Side Channel Attacks via Enclave Verifiable
                  Page Faults}},
  booktitle    = raid,
  year         = {2017},
}

@inproceedings{DBLP:conf/eurosys/ZhaoLQQ20,
  author       = {Wenjia Zhao and
                  Kangjie Lu and
                  Yong Qi and
                  Saiyu Qi},
  title        = {{MPTEE:} bringing flexible and efficient memory protection to Intel
                  {SGX}},
  booktitle    = eurosys,
  year         = {2020},
}

@inproceedings{DBLP:conf/ccs/BuhrenWS19,
  author       = {Robert Buhren and
                  Christian Werling and
                  Jean{-}Pierre Seifert},
  title        = {Insecure Until Proven Updated: Analyzing {AMD} SEV's Remote Attestation},
  booktitle    = acm-ccs,
  year         = {2019},
}

@misc{google_2024_arm_tau,
  author       = {Google},
  title        = {{Tau VM: the first Google Compute Engine VM running on an ARM chip}},
  year         = {2022},
  howpublished          = {\url{https://cloud.google.com/blog/products/compute/tau-t2a-is-first-compute-engine-vm-on-an-arm-chip}},
    note = {Accessed: 2025-08-25}
}

@misc{aws_2022_arm,
  author       = {{Amazon Web Services}},
  title        = {{AWS \& ARM Partnership}},
  year         = {2022},
  howpublished          = {\url{https://www.arm.com/partners/aws}},
    note = {Accessed: 2025-08-25}
}

@misc{azure_2022_arm,
  author       = {Microsoft},
  title        = {{Azure Virtual Machines with Ampere Altra ARM-based processors generally available}},
  year         = {2022},
  howpublished          = {\url{https://azure.microsoft.com/en-us/blog/azure-virtual-machines-with-ampere-altra-arm-based-processors-generally-available/}},
    note = {Accessed: 2025-08-25}
}

@misc{hyperenclave_git,
    title = {{HyperEnclave}},
    howpublished = {\url{https://github.com/asterinas/hyperenclave}},
    note = {Accessed: 2025-08-25}
}

@misc{linux_svm,
    title = {{Linux SVSM (Secure VM Service Module)}},
    howpublished = {\url{https://github.com/AMDESE/linux-svsm}},
    note = {Accessed: 2025-08-25}
}

@inproceedings{DBLP:conf/eurosys/TsaiABJJJKKOP14,
  author       = {Chia{-}Che Tsai and
                  Kumar Saurabh Arora and
                  Nehal Bandi and
                  Bhushan Jain and
                  William Jannen and
                  Jitin John and
                  Harry A. Kalodner and
                  Vrushali Kulkarni and
                  Daniela Oliveira and
                  Donald E. Porter},
  title        = {{Cooperation and security isolation of library OSes for multi-process
                  applications}},
  booktitle    = eurosys,
  year         = {2014},
}

@inproceedings{DBLP:conf/usenix/TsaiPV17,
  author       = {Chia{-}Che Tsai and
                  Donald E. Porter and
                  Mona Vij},
  title        = {{Graphene-SGX: {A} Practical Library {OS} for Unmodified Applications
                  on {SGX}}},
  booktitle    = usenix-atc,  
  year         = {2017},
}

@inproceedings{DBLP:conf/asplos/ShenTCCWXXY20,
  author       = {Youren Shen and
                  Hongliang Tian and
                  Yu Chen and
                  Kang Chen and
                  Runji Wang and
                  Yi Xu and
                  Yubin Xia and
                  Shoumeng Yan},
  title        = {{Occlum: Secure and Efficient Multitasking Inside a Single Enclave of Intel {SGX}}},
  booktitle    = asplos,
  year         = {2020},
}

@string{acm-csur = "ACM Computing Surveys (CSUR)" }

@string{acm-pomacs                = "ACM on Measurement and Analysis of Computing Systems (POMACS)"}

@string{acm-ccs         = "ACM Conference on Computer and Communications Security (CCS)"}

@string{usenix-security = "USENIX Security Symposium"}

@string{isoc-ndss       = "Symposium on Network and Distributed System Security (NDSS)"}

@string{raid            = "Symposium on Recent Advances in Intrusion Detection (RAID)"}

@string{esorics         = "European Symposium on Research in Computer Security (ESORICS)"}

@string{usenix-woot     = "USENIX Workshop on Offensive Technologies (WOOT)"}

@string{usenix-atc      = "USENIX Annual Technical Conference (ATC)"}

@string{ieee-infocom-mc = "IEEE Conference on Computer Communications (INFOCOM)"}

@string{dimva           = "Detection of Intrusions and Malware, and Vulnerability Assessment (DIMVA)"}

@string{oopsla          = "ACM SIGPLAN Conference on Object-Oriented Programming Systems, Languages, and Applications (OOPSLA)"}

@string{sas             = "International Static Analysis Symposium (SAS)"}

@string{asplos          = "Conference on Architectural Support for Programming Languages and Operating Systems (ASPLOS)"}

@string{eurosec         = "ACM European Workshop on System Security (EuroSec)"}

@string{eurosys         = "European Conference on Computer Systems (EuroSys)"}

@string{osdi            = "Symposium on Operating Systems Design and Implementation (OSDI)"}

@string{asia-ccs        = "ACM Symposium on Information, Computer and Communications Security (ASIACCS)"}

@string{ieee-sp         = "IEEE Symposium on Security and Privacy (S\&P)"}

@string{ieee-eurospw     = "IEEE European Symposium on Security and Privacy Workshops (EuroS\&PW)"}

@string{spw             = {{IEEE} Security and Privacy Workshops (SPW)}}

@string{seed = "International Symposium on Secure and Private Execution Environment Design (SEED)"}

@string{sas = "Static Analysis Symposium"}

@string{hasp = "International Workshop on Hardware and Architectural Support for Security and Privacy (HASP)"}

@string{sosp = "Symposium on Operating Systems Principles (SOSP)"}

@string{mobisys = "ACM International Conference on Mobile Systems, Applications, and Services (MobiSys)"}

@string{ches = "Conference on Cryptographic Hardware and Embedded Systems (CHES)"}

@string{systex = "Workshop on System Software for
Trusted Execution (SysTEX)"}

\newpage
\nobalance
\section*{Appendix}

In this appendix, we provide additional information about some of the contents of this paper. 
First, Table~\ref{tab:sgx_instructions} presents a detailed list of all the SGX microprograms available on modern Intel CPUs. 
Next, Table~\ref{tab:microinstr} shows more detailed numbers on the results of the microbenchmarks executed as part of the evaluation presented in Section~\ref{sec:eval}.

\begin{table*}[b]
\caption{
The following table presents an overview of the SGX \texttt{ENCLU} and \texttt{ENCLS} microinstructions. 
  The leaf-number in the \texttt{EAX} register decides the microprograms to be run. 
A short description for each program is given in the last column.
}
    \centering
    \resizebox{0.8\linewidth}{!}{
\begin{tabularx}{\linewidth}{cccX}
  \hline \Tstrut \textbf{Instruction}  & \textbf{EAX}  & \textbf{SGX Function} & \textbf{Description}                                    \Bstrut \\ \hline \Tstrut
  \textbf{ENCLS}        & 0x0           & ECREATE               & Create an SECS page in the EPC                          \\
                        & 0x1           & EADD                  & Add a Page to an Uninitialized Enclave                    \\
                        & 0x2           & EINIT                 & Initialize an Enclave for Execution                     \\
                        & 0x3           & EREMOVE               & Remove a page from the EPC                              \\
                        & 0x4           & EDGBRD                & Read From a Debug Enclave                               \\
                        & 0x5           & EDBGWR                & Write to a Debug Enclave                                \\
                        & 0x6           & EEXTEND               & Extend Uninitialized Enclave Measurement by 256 Bytes   \\
                        & 0x7           & ELDB                  & Load an EPC page and Marked its State Blocked           \\
                        & 0x8           & ELDU                  & Load an EPC page and Marked its State Unblocked         \\
                        & 0x9           & EBLOCK                & Mark a page in EPC as Blocked                           \\ 
                        & 0xA           & EPA                   & Add Version Array                                       \\
                        & 0xB           & EWB                   & Invalidate an EPC Page and Write out to Main Memory     \\ 
                        & 0xC           & ETRACK                & Activates EBLOCK Checks                                 \\ 
                        & 0xD           & EAUG                  & Add a Page to an Initialized Enclave                    \\ 
                        & 0xE           & EMODPR                & Restrict the Permissions of an EPC Page                 \\ 
                        & 0xF           & EMODT                 & Change the Type of an EPC Page                          \Bstrut\\ \hline \Tstrut 
  \textbf{ENCLU}        & 0x0           & EREPORT               & Create a Cryptographic Report of the Enclave            \\
                        & 0x1           & EGETKEY               & Retrieves a Cryptographic Key                           \\ 
                        & 0x2           & EENTER                & Enters an Enclave                                       \\
                        & 0x3           & ERESUME               & Re-Enters an Enclave                                    \\
                        & 0x4           & EEXTIT                & Exits an Enclave                                        \\ 
                        & 0x5           & EACCEPT               & Accept Changes to an EPC Page                           \\ 
                        & 0x6           & EMODPE                & Extend an EPC Page Permissions                          \\ 
                        & 0x7           & EACCEPTCOPY           & Initialize a Pending Page                               \\ 
                        & 0x9           & EDECCSSA              & Decrements TCS.CSSA                                     \\
 \hline
\end{tabularx}}
  \label{tab:sgx_instructions}
\end{table*}

\begin{table*}[b]
  \caption{Results of the microbenchmarks for the CCX microprograms with the results in nanoseconds. }
    \centering
    \resizebox{0.7\linewidth}{!}{
  \begin{tabularx}{\linewidth}{cX|cc}
  \hline 
  \Tstrut
  \textbf{Instruction} & \textbf{Runtime} & \textbf{Instruction} & \textbf{Runtime} \Bstrut \\ \hline \Tstrut
\textbf{ECREATE}    &15281521    &\textbf{EAUG} & 29040  \\
\textbf{EADD}       &18840    &\textbf{EMODPR} & 6360 \\
\textbf{EINIT}      &1898520  &\textbf{EMODT} & 3960 \\
\textbf{EREMOVE}    &3000     &\textbf{EREPORT} & 114639 \\
\textbf{EDGBRD}     &6000     &\textbf{EGETKEY} & 53440  \\
\textbf{EDBGWR}     &5880     &\textbf{EENTER} & 15153  \\
\textbf{EEXTEND}    &25440    &\textbf{ERESUME} & 13664 \\
\textbf{EDLU/B}     &843360   &\textbf{EEXIT} & 2334 \\
\textbf{EBLOCK}     &4320     &\textbf{EACCEPT} & 5773 \\
\textbf{EPA}        &24840    &\textbf{EMODPE} & 5802 \\
\textbf{EWB}        &910200 &\textbf{EACCEPTCOPY} & 32606 \\
\textbf{ETRACK}     &1920 &\textbf{EDECCSSA} & 3750 \Bstrut \\ \hline 
\end{tabularx}}
\label{tab:microinstr}
\end{table*}

\end{document}